\documentclass[aps,prd,twocolumn,preprintnumbers,superscriptaddress,
            nofootinbib]{revtex4}
\pdfoutput=1

\newcommand{\PRE}[1]{}				

\usepackage{amsmath}
\usepackage{amsfonts}
\usepackage{amssymb}
\usepackage{epsfig}
\usepackage{hyperref}
\usepackage{graphicx}
\usepackage{setspace}

\newcommand{\beq}{\begin{equation}}
\newcommand{\eeq}{\end{equation}}

\newcommand{\wjmL}{\left(
                         \begin{array}{ccc}
       l_1 & l_2  & L  \\
         m_1 & m_2  & M
                         \end{array}
                   \right)}
\newcommand{\wjmmL}{\left(
                         \begin{array}{ccc}
       l_3 & l_4  & L  \\
         m_3 & m_4  & -M
                         \end{array}
                   \right)}

\newcommand{\n}{{\bf{\hat n}}}

\def\myK{{\cal K}}

\newcommand{\wjjj}[6]
{{
\left(
\begin{array}{lcr} #1 & #2 & #3 \\#4 & #5 & #6 \end{array}
\right)
}}

\begin{document}

\title{\PRE{\vspace*{1.5in}}
CMB Constraints on Primordial non-Gaussianity from the Bispectrum ($f_{\rm NL}$) and Trispectrum ($g_{\rm NL}$ and $\tau_{\rm NL}$) and a New Consistency Test of Single-Field Inflation
\PRE{\vspace*{0.3in}}}

\author{Joseph Smidt\footnote{ jsmidt@uci.ed}}
\affiliation{Center for Cosmology, Department of Physics and Astronomy,
University of California, Irvine, CA 92697, USA }

\author{Alexandre Amblard\footnote{amblard@uci.edu}} 
\affiliation{Center for Cosmology, Department of Physics and Astronomy,
University of California, Irvine, CA 92697, USA }

\author{Christian T. Byrnes\footnote{Byrnes@physik.uni-bielefeld.de}} 
\affiliation{Fakult\"at f\"ur Physik, Universit\"at Bielefeld, Postfach 100131, 33501 Bielefeld, Germany }

\author{Asantha Cooray\footnote{acooray@uci.edu}} 
\affiliation{Center for Cosmology, Department of Physics and Astronomy,
University of California, Irvine, CA 92697, USA }

\author{Alan Heavens\footnote{afh@roe.ac.uk}} 
\affiliation{Scottish Universities Physics Alliance (SUPA),~ Institute for Astronomy,
University of Edinburgh, Blackford Hill,  Edinburgh EH9 3HJ, UK}

\author{Dipak Munshi\footnote{dipak.munshi@astro.cf.ac.uk}} 
\affiliation{Scottish Universities Physics Alliance (SUPA),~ Institute for Astronomy,
University of Edinburgh, Blackford Hill,  Edinburgh EH9 3HJ, UK}
\affiliation{School of Physics and Astronomy, Cardiff University, CF24 3AA}
\date{\today}

\begin{abstract}
\PRE{\vspace*{.3in}}
We outline the expected constraints on non-Gaussianity from the cosmic
microwave background (CMB) with current and future experiments, focusing on
both the third ($f_{\rm NL}$) and fourth-order  ($g_{\rm NL}$ and $\tau_{\rm
NL}$) amplitudes of the local configuration or non-Gaussianity.  The
experimental focus is the skewness (two-to-one) and kurtosis (two-to-two and
three-to-one) power spectra from weighted maps. In adition to a measurement of $\tau_{\rm NL}$ and $g_{\rm NL}$ with WMAP 5-year data, our study
provides the first forecasts for future constraints on $g_{\rm NL}$. We describe how these
statistics can be corrected for the mask and cut-sky through a window function,
bypassing the need to compute linear terms that were  introduced for the
previous-generation non-Gaussianity statistics, such as the skewness estimator.
We discus the ratio $A_{\rm NL} =  \tau_{\rm NL}/(6f_{\rm NL}/5)^2$ as an
additional test of single-field inflationary models and discuss the physical
significance of each statistic.  Using these estimators with WMAP 5-Year V+W-band data out to $l_{\rm max}=600$ we constrain  the cubic order
non-Gaussianity parameters $\tau_{\rm NL}$, and $g_{\rm NL}$ and find  $-7.4 < g_{\rm NL}/10^5 < 8.2$ and   $-0.6 < \tau_{\rm NL}/10^4 < 3.3$ improving the previous COBE-based limit on $\tau_{\rm NL} < 10^8$ nearly four orders of magnitude with WMAP. 

\end{abstract}


\maketitle


\section{Introduction}

We have now entered an exciting time in cosmological studies where we are now
beginning to constrain simple slow-roll inflationary models with high precision
observations of the cosmic microwave background (CMB) and large-scale
structure.  In addition to constraining inflationary model parameter space with
traditional parameters such as the spectral index $n_s$ and the
tensor-to-scalar ratio $r$, we may soon be able to use parameters associated
with primordial non-Gaussianity to improve model selection.

\begin{figure}
    \begin{center}
      \includegraphics[scale=0.25]{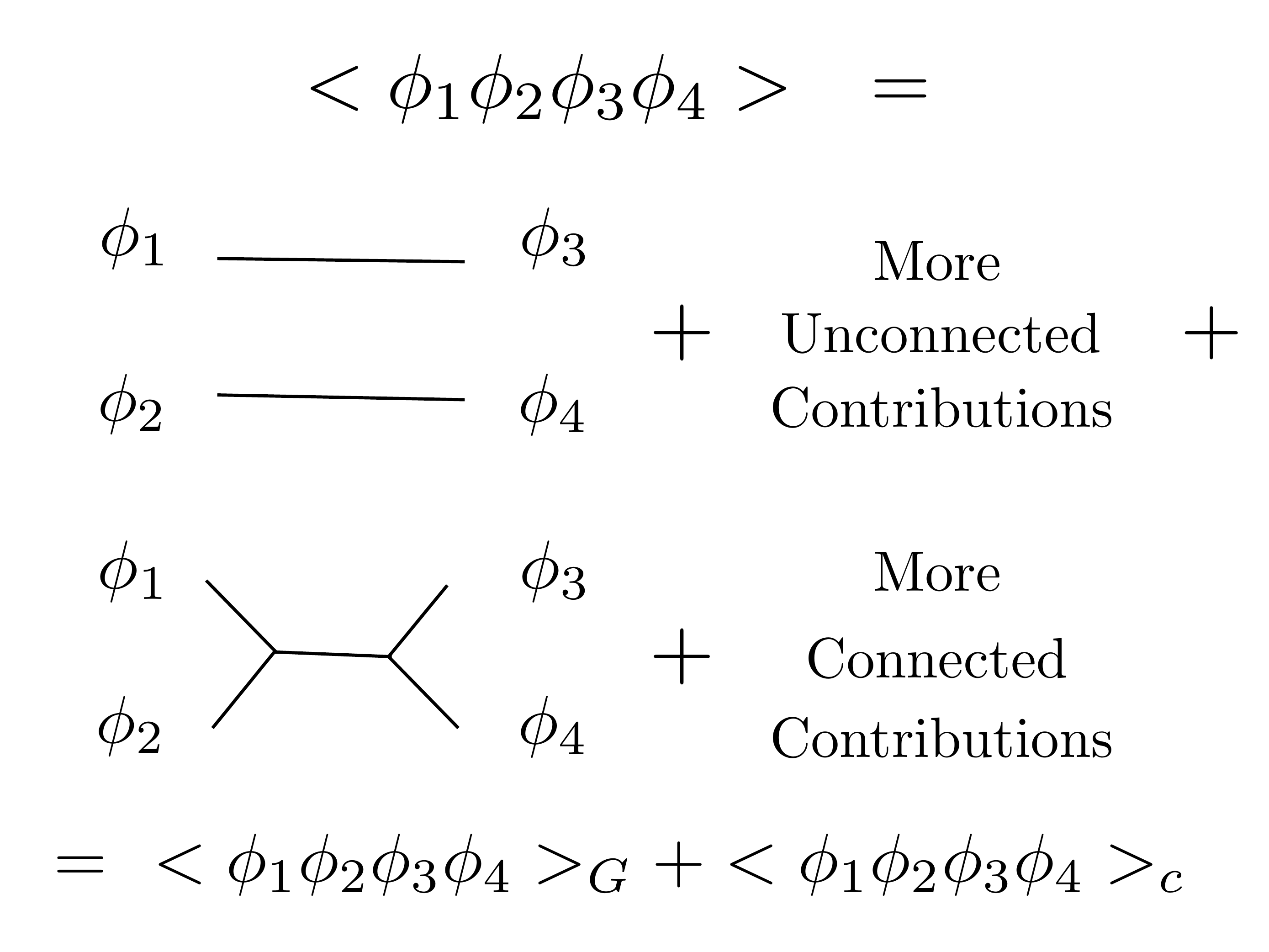} 
   \end{center}
   \vspace{-0.5cm}
   \caption[width=1in]{Four point correlation function for the $\phi^3$ theory.  The correlation functions breaks up into interaction-less unconnected diagrams and connected diagrams containing information about the interactions. }
   \label{fig:Feynman}
\end{figure}

In the simplest realistic inflationary models, the field(s) responsible for
inflation have minimal interactions.  Such an interaction-less situation should
have led to Gaussian primordial curvature perturbations, assuming that pertubations in the inflaton field generates the curvature perturbation. In this case, the two
point correlation function contains all the informations on these
perturbations.  If the early inflation field(s) have non-trivial interactions,
higher-order correlation functions  of the curvature perturbations will contain
{\it connected } pieces encoding information about the primordial inflationary
interactions.  This is analogous to the situation encountered in particle
physics where correlation functions can be separated into unconnected and
connected Feynman diagrams, the later containing information about the
underlying interactions (see Fig.~\ref{fig:Feynman} for an example involving
the four-point function).  A detection of non-Gaussianity therefore gives an
important window into the nature of the inflation field(s) and their
interactions.

To parameterize the non-Gaussianity of a nearly Gaussian field, such as the
primordial curvature perturbations $\zeta({\bf x})$,  we can expand them
perturbatively~\cite{Kogo:2006kh} to second order as:
\begin{equation}
\zeta({\bf x}) = \zeta_g({\bf x}) + {3 \over 5}f_{\rm NL} \left[\zeta_g^2({\bf x}) - \langle \zeta^2_g({\bf x})\rangle\right] + {9 \over 25 } g_{\rm NL} \zeta_g^3({\bf x}), 
\label{eq:phi}
\end{equation}
 where $\zeta_g({\bf x})$ is the purely Gaussian part with $f_{\rm NL}$ and
$g_{\rm NL}$ parametrizing  the first and second order deviations from
Gaussianity.   This parameterization of the curvature perturbations is known as
the local model as this definition is local in space.

Much effort has already gone into measuring non-Gaussianity at first-order in
curvature perturbations using the bispectrum of the CMB anisotropies or
large-scale structure galaxy distribution parametrerized by $f_{\rm NL}$ (see
Eq.~\ref{eq:phi}).  These studies have found $f_{\rm NL}$ to be consistent with
zero~\cite{Yadav:2007yy,Smith:2009jr,Komatsu:2010fb,Smidt:2009ir}.  However,
there is hope that a significant detection may be possible by future surveys
that will lead to improved errors~\cite{Komatsu:2009kd}.  

In the trispectrum, two parameters of second-order non-Gaussianity at
fourth-order in curvature perturbations, $\tau_{\rm NL}$ and $g_{\rm NL}$, can
be measured.  In this paper we also introduce a third parameter, $A_{\rm NL}$
is an additional parameter that compares $\tau_{\rm NL}$ of the trispectrum to
$(6f_{\rm NL}/5)^2$ from the bispectrum as a ratio:
\begin{equation}
A_{\rm NL} = {\tau_{\rm NL} \over (6 f_{\rm NL}/ 5)^2}.
\end{equation}
This ratio can be quite different for many inflationary
models~\cite{Byrnes:2010em,Chen:2009bc} and, as will be shown
below, $A_{\rm NL} \neq 1$ rules out single-field inflationary models
altogether, including the standard curvaton scenario (which neglects perturbations from the inflaton field). 
 
 In this paper we discuss the skewness and kurtosis power spectra method for probing
primordial non-Gaussianity and give constraints for the first ($f_{\rm NL}$)
and second-order ($g_{\rm NL}$ and $\tau_{\rm NL}$) amplitudes of the local
model in addition to their ratio $A_{\rm NL}$.  Using the bispectrum of CMB
anisotropies as seen by WMAP 5-year data, Smidt et al.~(2009) found $-36.4 <
f_{\rm NL} < 58.4$ at 95\% confidence~\cite{Smidt:2009ir}. This is to be
compared with the most recent WMAP 7 measurement of $-10 < f_{\rm NL} <
74$~\cite{Komatsu:2010fb}, where part of the discrepancy is due to a difference
in optimization~\cite{Smith}.   As outlined in Section~VI, using the trispectrum of the same data we find that  $-0.6 < \tau_{\rm NL}/10^4 < 3.3$ and $-7.4 < g_{\rm
NL}/10^5 < 8.2$ at 95\% confidence level showing second order non-Gaussianity
is consistent with zero in WMAP.  This paper serves as a guide to the analysis process behind our derived limits on $\tau_{\rm NL}$,
$g_{\rm NL}$ and $A_{\rm NL}$.
 
Furthermore, in this paper we analyze what to realistically expect when
measuring non-Gaussianity from CMB temperature data.  We believe establishing
what constraints can be placed upon $f_{\rm NL}$, $\tau_{\rm NL}$, $g_{\rm NL}$
and $A_{\rm NL}$ by future experiments is important in determining what models
may and may not be tested by future data.  We also highlight several advantages
of our work, including ways to correct the cut-sky and mask through a window
function without using linear terms which are computationally
prohibative~\cite{Creminelli:2005hu,Yadav:2007ny}.

This paper is organized as follows:  In Section~\ref{sec:Inf} we review how
non-Gaussianity may be used to distinguish between common inflationary models
and stress the physical significance of each statistic. In
Section~\ref{sec:Theory} we describe the skewness and kurtosis power spectra
and explain how they may be used to extract information about primordial
non-Gaussianity from the CMB.  In Section~\ref{sec:fishertheory}, we describe
the signal-to-noise of each estimator, how to add the experimental beam and
noise to these calculations and discuss why these power spectra have the
advantage for dealing with a cut sky.   In Section~\ref{sec:fisherresults} we
calculate the fisher bounds for upcoming experiments for each statistic. In Section~\ref{sec:prioranalysis} we discuss the technical details for measuring non-Gaussianity in the trispectrum  and in
Section~\ref{sec:conclusion} we conclude with a discussion.

%
%

\section{Non-Gaussianity From Common Inflationary Models}
\label{sec:Inf}

Non-Gaussinity is a powerful tool that may be used to distinguish between
inflationary models.  The simplest models do not produce a detectable amount of
non-Gaussianity.  Maldacena~\cite{Maldacena:2002vr} has shown that a
single-field, experiencing slow roll with canonical kinetic energy and an
initial Bunch-Davies vacuum state produces

\begin{equation}
\label{eq:f2ns}
f_{\rm NL} = {5 \over 12} (n_s + f(k) n_t).
\end{equation}
Here $n_s$ and $n_t$ are the scalar and tensor spectral indices respectively.
The function  $f(k)$ has a range $0 \le f(k) \le {5 \over 6}$ based on the
triangle shapes (see below) of the $k_i$ such that $f = 0$ in the squeezed
limit and $f = {5 \over 6}$ for an equilateral triangle.  For this reason,
$f_{\rm NL} < 1$ will remain undetectable in the simple slow roll scenario with
CMB data alone.
\begin{figure}
    \begin{center}
      {\includegraphics[scale=0.45]{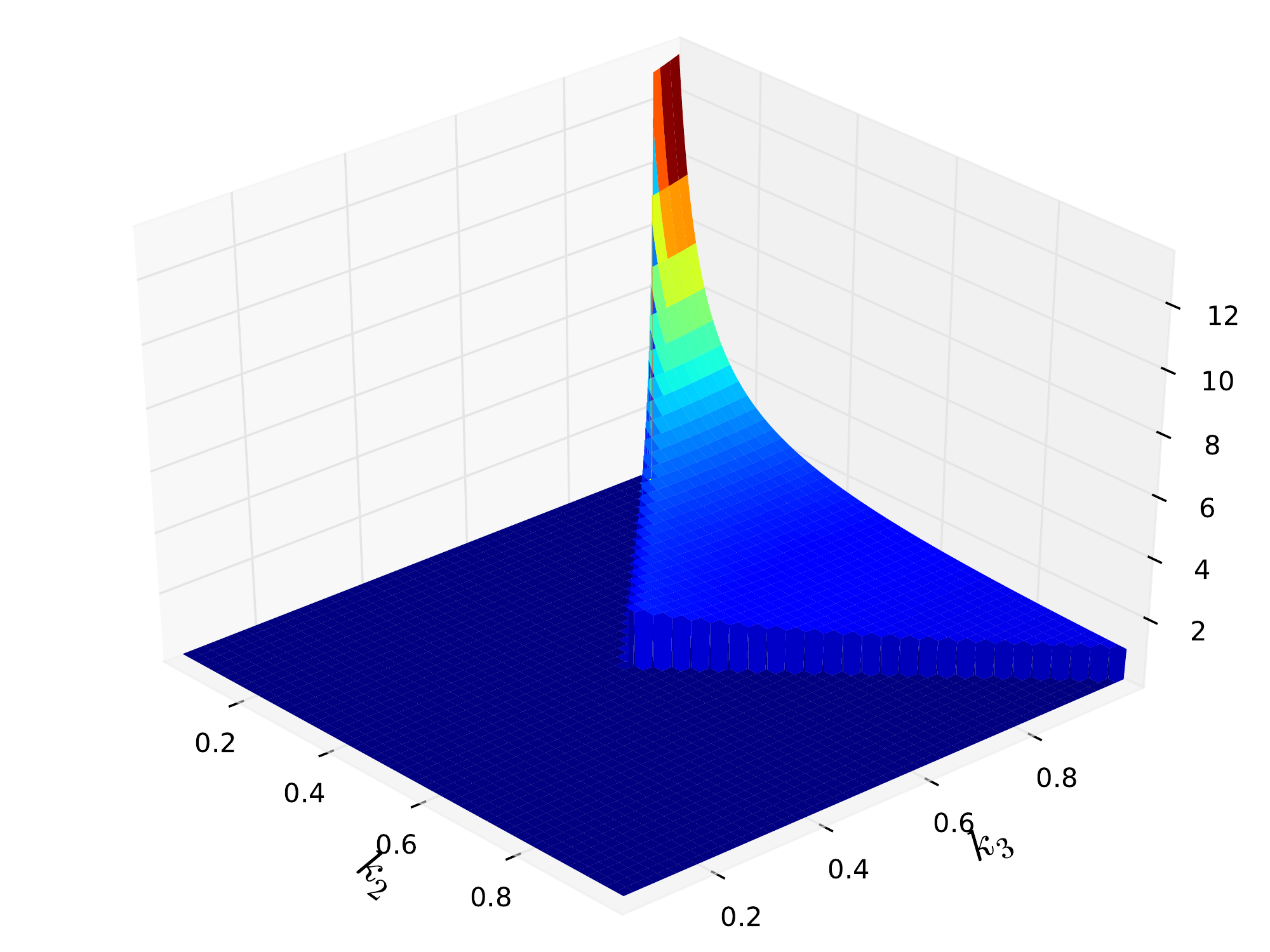} 
       \includegraphics[scale=0.45]{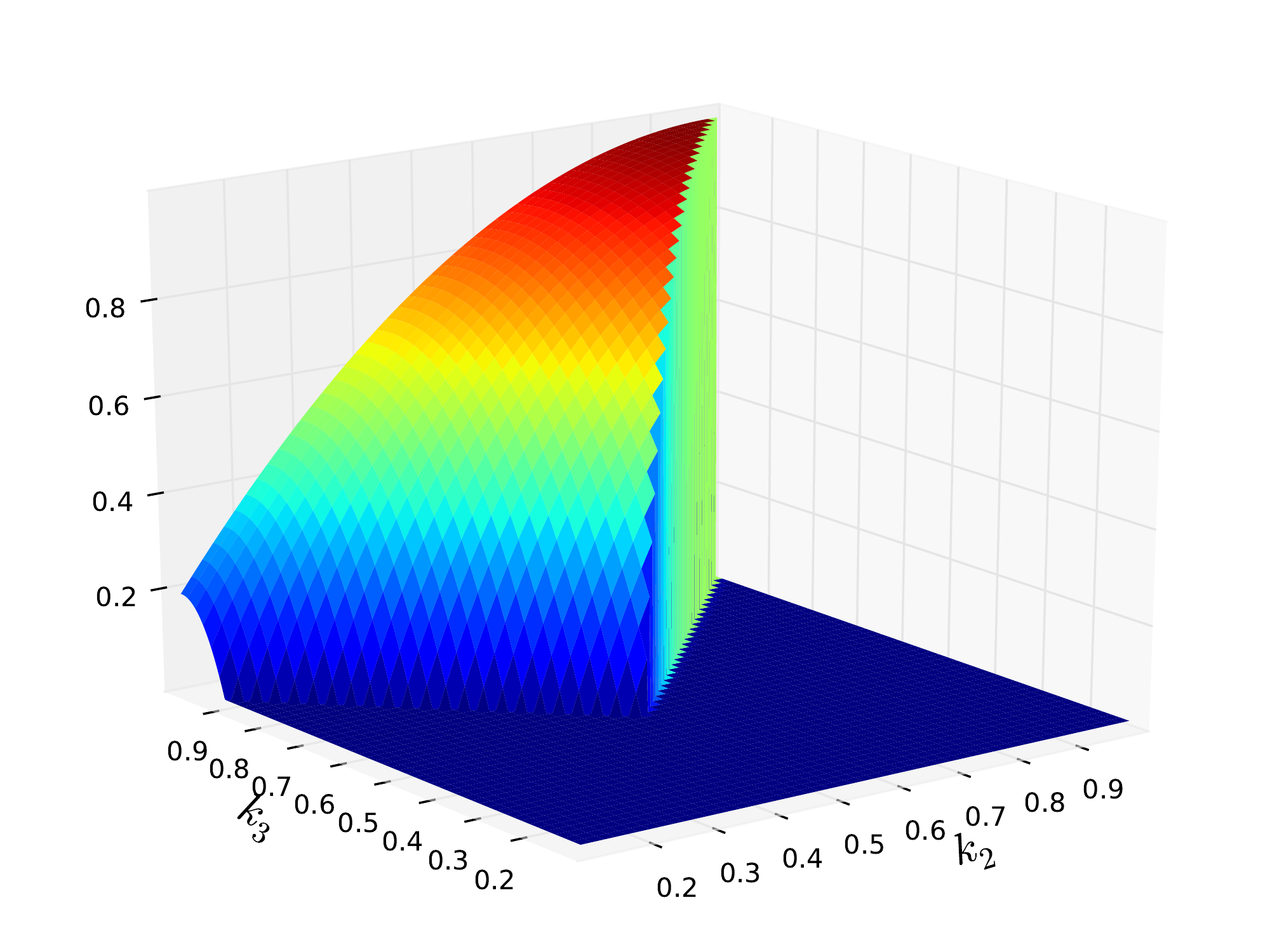} }
   \end{center}
   \vspace{-0.7cm}
   \caption[width=3in]{Plot of the shape functions $S^{\rm local}(1, k_2, k_3)$ and $S^{\rm equil}(1, k_2, k_3)$ normalized such that $S(1, 1, 1) = 1$.  In these plots only values satisfying the triangle inequality $k_2 + k_3 \geq k_1 = 1$ as well as the requirement $k_2 \leq k_3$ to prevent showing equivalent configurations are non-zero. The plot on top verified $S^{local}$ is maximized when $k_1 \sim k_3 \gg k_2$ whereas the bottom plot verifies $S^{equal}$ is maximised when $k_1 \sim k_2 \sim k_3$.}
   \label{fig:shapes}
\end{figure}
If any of the above assumptions are violated, very specific types of
non-Gaussianity are produced~\cite{Komatsu:2009kd,Baumann:2009ds,
Komatsu:2010hc}. In the bispectrum $B_\zeta( k_1, k_2, k_3)$ defined by

\begin{equation}
\left<\zeta_{{\bf k}_1} \zeta_{{\bf k}_2} \zeta_{{\bf k}_3}\right>  = (2 \pi)^3 \delta({\bf k}_1 + {\bf k}_2 + {\bf k}_3) B_\zeta(k_1,  k_2, k_3),
\end{equation}
where $\zeta$ is the primordial curvature perturbation, non-Gaussianities show
up as triangles in Fourier space.   Different triangle shapes are be produced
by different underlying physics, for example:
\begin{itemize}
\item {\it squeezed triangle} ($k_1 \sim k_2 \gg k_3$) This is the dominating shape from multi-field, curvaton, inhomogeneous reheating and Ekpyrotic models.
\item {\it equilateral  triangle} ($k_1 = k_2 = k_3$)  This shape is produced by non-canonical kinetic energy with higher derivative interactions and non-trivial speeds of sound.
\item {\it folded triangle} ($k_1 = 2 k_2 = 2 k_3$)  These triangles are produced by non-adiabatic-vacuum models. 
\end{itemize}
Additionally, linear combinations of the above shapes or intermediate cases
such as {\it elongated triangles} ($k_1 = k_2 + k_3$) and {\it isosceles
triangles} ($k_1 > k_2 = k_3$) are
possible~\cite{Komatsu:2009kd,Baumann:2009ds, Komatsu:2010hc}.  The most recent
WMAP 7 constraints on the amount of non-Gaussinaity from each shape is $-10 <
f^{\rm local}_{\rm NL} < 74$, $-214 < f^{\rm equil}_{\rm NL} < 266$ and $-410 <
f^{\rm orthog}_{\rm NL} < 6$ at 95\% confidence~\cite{Komatsu:2010fb}.

A convenient way to distinguish between shapes is to introduce the shape
function defined as 
\begin{equation}
S(k_1, k_2, k_3) \equiv {1 \over N}(k_1k_2 k_3)^2 B_\zeta(k_1, k_2, k_3),
\end{equation}
where N is a normalization factor often taken to be $1/f_{\rm NL}$.  Using a
notation introduced by Fergusson and Shellard~\cite{Fergusson}, we can give the
shape function for the more common configurations as:
\begin{eqnarray}
S^{\rm local}(k_1, k_2, k_3) &\propto& {K_3 \over K_{111}}, \\
S^{\rm equil}(k_1, k_2, k_3)&\propto& {\tilde k_1 \tilde k_2 \tilde k_3 \over K_{111}},\\
S^{\rm folded}(k_1, k_2, k_3) &\propto& {1 \over K_{111}}(K_{12}-K_3)+4{K_2 \over \tilde k_1 \tilde k_2 \tilde k_3 }, 
\end{eqnarray} 
where
\begin{eqnarray}
K_p &=& \sum_i (k_i)^p \ \ \ {\rm with} \ \ \ K = K_1,\\
K_{pq} &=& {1 \over \Delta_{pq}} \sum_{i \neq j} (k_i)^p(k_j)^q,\\
K_{pqr} &=& {1 \over \Delta_{pqr}} \sum_{i \neq j \neq l} (k_i)^p (k_j)^q (k_l)^r,\\
\tilde k_{ip} &=& K_p - 2(k_i)^p \ \ \  {\rm with} \ \ \ \tilde k_i = \tilde k_{i 1},
\end{eqnarray} 
with  $\Delta_{pq} = 1 + \delta_{pq}$ and $\Delta_{pqr} =
\Delta_{pq}(\Delta_{qr}+\delta_{pr})$ (no summation). Plots for the local and
equilateral shapes are given in Figure~\ref{fig:shapes}.

In addition to $f_{\rm NL}$ being generated by different shapes, it also may
vary with scale.  Recently, a new parameter has been introduced to measure this
scale dependance defined as:
\begin{equation}
n_{f_{\rm NL}}(k) = {d \ln{ \left|f_{\rm NL}(k)\right|} \over d \ln k}.
\end{equation}
This scale dependance has the ability to test the ansatz~\ref{eq:phi} to test
whether the local model should allow for $f_{\rm NL}$ to vary with
scale~\cite{Byrnes:2009pe}.  Using the results of Smidt et al.(2009) (Fig. 16
of Ref~\cite{Smidt:2009ir}) and assuming 
\begin{equation}
\label{eq:nfnll}
f_{\rm NL}(l) = f_{\rm NL_{200}} \left({l \over l_{200}}\right)^{n_{f_{\rm NL}}(l)},
\end{equation}
we can constrain $n_{f_{\rm NL}}(l)$ to roughly $-2.5 < n_{f_{\rm NL}}(l) < 2.3$ at
95\% confidence.  We therefore find $f_{\rm NL}$ is consistent with having no
scale dependance.

In this paper we focus on the local model that probes non-Gaussianty of a
squeezed shape.  As mentioned above, simple inflationary models can not produce
a detectable amount of non-Gaussinity for local models.  We now review the
prediction for local non-Gaussianity for the most common models.  

%
%

\subsection{Review Of The $\delta N$ formalism.}

The curvature perturbation can be conveniently described using the $\delta N$ formalism~\cite{Starobinsky:1982ee,Starobinsky:1986fxa,Byrnes:2006vq, Lyth:2005fi, Suyama:2007bg}.  During inflation, spacetime expands by a certain number of e-folds N.   By Heisenberg's uncertainty principle, expansion for each point in space ends at slightly different times producing a spatially dependent total e-fold:
\begin{equation}
N(x) = \int_{t_i}^{t^f} H(x,t) dt,
\end{equation}
where $H(x,t)$ is the Hubble parameter allowing us to define $N(x) = {\bar N} +
\delta N(x)$.  The fluctuations in e-fold about the mean value ${\bar N}$,
which correspond to perturbations in local expansion, are the curvature
perturbations $\zeta = \delta N$.

In addition to a spatial parameterization, we may parameterize the number of
e-folds by the underlying fields $\zeta = N(\phi^A) - {\bar N}$ where $\phi^A$
represents the initial values for the scalar fields .  If we write out the
fields as $\phi^A = {\bar \phi}^A + \delta \phi^A$ we can expand the curvature
perturbations as
\begin{equation}
\zeta = \delta N = \sum_n {1 \over n!} N_{A_1 A_2 ... A_n} \delta \varphi^{A_1} \delta \varphi^{A_1} ... \delta \varphi^{A_n}.
\end{equation}
The $N_x$ means the derivative of N with respect to the fields $x$.  For
example, $N_{A_1 A_2} \equiv {\partial^2 N \over \partial \varphi^{A_1}
\partial \varphi^{A_2}}$.  In this equation there is an implicit sum over the
$A_i$.  Einstein summation is implicit in all equations relating to the $\delta
N$ formalism.

Using this formalism we may compute to first order from $\zeta = N_{ A} \delta
\varphi^{A}$:
\begin{eqnarray}
 \left< \zeta_{\bf k} \zeta_{\bf k'}\right> &=& N_{ A} N_{ B} C^{A B} (k) (2 \pi)^3 \delta^3({\bf k} + {\bf k'}), 
\end{eqnarray}
where $C^{A B}(k)$ in the slow roll limit becomes to leading order $\delta^{A
B} P(k)$.

Likewise, we can calculate the bispectrum and trispectrum in this formalism;
\begin{eqnarray}
B_{\zeta}(k_1, k_2, k_3) &=& \hspace{-0.1 cm}N_A N_{B C} N_D [C^{AB}(k_1) C^{BD}(k_2)  \\
&& \hspace{-1cm}+ \ C^{AB}(k_1) C^{BD}(k_2) + C^{AB}(k_1) C^{BD}(k_2)],  \nonumber \\
&& \nonumber \\
T_\zeta(k_1, k_2, k_3, k_4) &=& N_{A_1 A_2}N_{B_1 B_2}N_CN_D \\
  \times [ C^{A_2 B_2}(k_{13}) && \hspace{-0.7cm}C^{A_1C}(k_2)  C^{B_1 D}(k_2) + (11\; {\rm perms)}] \nonumber \\
  && + N_{A_1 A_2 A_3}N_{B}N_CN_D \nonumber \\
  \times [ C^{A_1 B}(k_{13}) && \hspace{-0.7cm}C^{A_2C}(k_2)  C^{A_3 D}(k_2) + (3 {\rm \; perms)}], \nonumber
\end{eqnarray}
where $k_{i j} = | {\bf k}_i + {\bf k}_j |$. In the slow roll limit to leading
order these expressions may be rewritten as:
\begin{eqnarray}
B_\zeta(k_1, k_2, k_3) &=& {6\over 5} f_{\rm NL} [P_\zeta(k_1) P_\zeta(k_2) + \\
&& P_\zeta(k_2) P_\zeta(k_3) + P_\zeta(k_3) P_\zeta(k_1)],  \nonumber \\
&& \nonumber \\
T(k_1, k_2, k_3, k_4) &=& \tau_{\rm NL} [ P_\zeta(k_{13}) P_\zeta(k_3) P_\zeta(k_4)  + (11\; {\rm perms)}] \nonumber \\
 &&  \hspace{-1.2 cm} + \ {54\over25} g_{\rm NL} [ P_\zeta(k_{2}) P_\zeta(k_3) P_\zeta(k_4) + (3\; {\rm perms)}], \label{taug}
\end{eqnarray}
where $P_\zeta(k) = N_A N_B C^{AB}(k)$ and therefore in the slow roll limit
$P_\zeta(k) = N_A N^A P(k)$.

From the above two expressions we can read off the values for each statistic:
\begin{eqnarray}
\label{eq:Nfnl}
f_{\rm NL} &=& {5\over 6} {N_A N_B N^{AB} \over \left( N_C N^C\right)^2};\\
\label{eq:Ntnl}
\tau_{\rm NL} &=& {N_{AB} N^{AC}N^B N_{C} \over \left( N_D N^D\right)^3};\\
\label{eq:Ngnl}
g_{\rm NL} &=& {25 \over 54} {N_{ABC} N^{A}N^B N^{C} \over \left( N_D N^D\right)^3}; \\
A_{\rm NL} &=&{ \tau_{\rm NL} \over (6 f_{\rm NL}/5)^2}.
\end{eqnarray}
%

%
%

\subsection{General Single-Field Models}

For a single scalar field $\varphi$ perturbing $N(\varphi)$ we may expand
$\zeta$, using the above formalism~\cite{Byrnes:2006vq}, as:
\begin{equation}
\zeta = N^\prime \delta \varphi + {1 \over 2} N^{\prime \prime} \delta \varphi^2 + {1 \over 6} N^{\prime \prime \prime} \delta \varphi^3 + ...,
\end{equation}
where $N^\prime = dN/d\varphi$. Note that we do not require that $\varphi$ is the inflaton field, it could be the curvaton or a field which modulates the efficiency of reheating. From equations~\ref{eq:Nfnl}-\ref{eq:Ngnl} we
may immediately read off 
\begin{eqnarray}
\label{eq:sfnl}
f_{\rm NL} &=& {5 \over 6} {N^{\prime \prime} \over \left(N^\prime \right)^2};\\
\label{eq:stnl}
\tau_{\rm NL} &=& {\left(N^{\prime \prime} \right)^2 \over \left(N^\prime \right)^4};\\
\label{eq:sgnl}
g_{\rm NL} &=& {25 \over 54} {N^{\prime \prime \prime} \over \left(N^\prime \right)^3};\\
A_{\rm NL} &=& 1.
\end{eqnarray}

Equations~\ref{eq:sfnl} and~\ref{eq:stnl} yield a very important consequence of
single-field models namely  $\tau_{\rm NL} = (6 f_{\rm NL}/5)^2$.  This is a
general result and therefore $A_{\rm NL} \neq 1$ may be used to rule out
single-field models all together.

%
%

\subsection{Multi-Field Inflationary Models}

Suyama and Yamaguchi showed in general $\tau_{\rm NL} \ge (6 f_{\rm NL}/5)^2$
by the Cauchy-Schwartz inequality and equality only holds if $N_{A}$ is an
eigenmode of $N_{AB}$~\cite{Suyama:2007bg}.    Models where equality does not hold can not
be those of a single-field.  We now examine such models.

Unlike the single-field case, using the $\delta N$ formalism to make general
statements about multi-field models is nearly impossible.  Instead, one is
forced to work with specific models that utilize simplifying assumptions.  We
now present a class of multi-field models that we believe is sufficiently
general to uncover many details that are characteristic of multi-field models in
general.

Recently, Byrnes and Choi reviewed two field models with scalar fields
$\varphi$ and $\chi$ that have a separable potential $W(\varphi, \chi) =
U(\varphi) V(\chi)$~\cite{Byrnes:2010em, Vernizzi:2006ve,Choi:2007su,
Battefeld:2006sz, Seery:2006js}.    The slow roll parameters for these models
are:
\begin{eqnarray}
\epsilon_\varphi &=& {M_p^2 \over 2} \left( {U_{, \varphi} \over U}  \right)^2,\; \epsilon_\chi = {M_p^2 \over 2} \left( {V_{, \chi} \over V}  \right)^2, \\ 
\\
\eta_{\varphi \varphi} \hspace{-0.1 cm}&=& \hspace{-0.1 cm}M_p^2 {U_{, \varphi \varphi} \over U}, \hspace{0.1 cm} \eta_{\varphi \chi} =  \hspace{-0.1 cm} M_p^2 {U_{, \varphi} V_{, \chi} \over W},  \hspace{0.1 cm} \eta_{\chi \chi} =  \hspace{-0.1 cm} M_p^2 {V_{, \chi \chi} \over V},   \nonumber
\end{eqnarray}
from which we can define
\begin{equation}
\tilde r = {\epsilon_{\chi} \over \epsilon_\varphi} e^{2(\eta_{\varphi \varphi} - \eta_{\chi \chi})N}.
\end{equation}

For this class of models, in the regions where $|f_{NL}|>1$ we have
\begin{eqnarray}
f_{\rm NL} &=& {5 \over 6} \eta_{\chi \chi}  {\tilde r \over (1 + \tilde r)^2} e^{2(\eta_{\varphi \varphi} - \eta_{\chi \chi})N};\\
g_{\rm NL} &=& {10 \over 3}  {\tilde r (\eta_{\varphi \varphi} - 2 \eta_{\chi \chi}) - \eta_{\chi \chi} \over 1 + \tilde r}  f_{\rm NL};\\
\tau_{\rm NL} &=& {1 + \tilde r \over \tilde r} \left({6 f_{\rm NL} \over 5}\right)^2; \\
A_{\rm NL} &=& {1 + \tilde r \over \tilde r}.
\end{eqnarray}

It is worth noting that both $\tau_{\rm NL}$ and $g_{\rm NL}$ are related to
$f_{\rm NL}$ for this class of models.  Here we have $|g_{\rm NL}| < |f_{\rm NL}|$
which will therefore be much harder to detect.  On the contrary, $\tau_{\rm NL} >
(6 f_{\rm NL}/5)^2$ so that non-Gaussinity may in fact  be easier to detect in
the trispectrum than the bispectrum for some multi-field models.  Here we find
$A_{\rm NL} =  {(1 + \tilde r) / \tilde r} > 1$.  The scale dependance of
$f_{\rm NL}$ has also been worked out for this class of models and was found to
be  $n_{f_{\rm NL}}=-4(\eta_{\varphi \varphi} - \eta_{\chi \chi})/(1+\tilde r)  < 0$.

%
%

\subsection{Curvaton Models}

In the curvaton scenario, a weakly interacting scalar field $\chi$ exists in
conjunction to the inflaton $\varphi$~\cite{Enqvist:2009N,Enqvist:2009T,Byrnes:2010em,
Byrnes:2006vq,Huang:2008bg,Enqvist:2009ww}.  During inflation, the curvaton
field is subdominant, but after inflation $\chi$ can dominate the energy
density.  The decay of the inhomogeneous curvation field in this scenario
produces the curvature perturbations and not the inflaton.

If such a curvaton field is the soul contributor to curvature perturbations, we
can write out the perturbations using the $\delta N$ formalism as we did in the
single field case:
\begin{equation}
\label{eq:curveN}
\zeta = N^\prime \delta \chi + {1 \over 2} N^{\prime \prime} \delta \chi^2 + {1 \over 6} N^{\prime \prime \prime} \delta \chi^3 + ...,
\end{equation}
where now $N^\prime = dN/d\chi$.  Immediately we recover the
relations~\ref{eq:sfnl}-\ref{eq:sgnl} and find for such curvaton models $A_{\rm
NL} = 1$ as should be expected from curvature perturbations generated by a
single-field. 

Recently, curvation models with generic potentials of the form
\begin{equation}
V = {1 \over 2} m^2 \chi^2 + \lambda \chi^{n+4},
\end{equation}
have been analyzed~\cite{Huang:2008bg,Enqvist:2009ww}.  Here $m$ is the
curvaton's mass and $\lambda$ is a coupling constant.  For such models $N$ in
Equation~\ref{eq:curveN} has been worked out giving:
\begin{eqnarray}
f_{\rm NL} &=& {5 \over 4 r_\chi} (1 + h) - {5 \over 3} - {5 r_\chi \over 6}, \\
g_{\rm NL} &=& {25 \over 54} \Bigg[  {9 \over 4 r_\chi^2}(\tilde h + 3h) - {9 \over r_\chi} (1+h) \\
&& + {1 \over 2}(1-9h) + 10 r_\chi + 3 r_\chi^2\Bigg], \nonumber
\end{eqnarray}
where
\begin{eqnarray}
r_{\chi} = {3 \Omega_{\chi, D} \over 4 - \Omega_{\chi, D} }, 
\hspace{0.3 cm} h = {\chi_0 \chi_0^{\prime \prime} \over  \chi_0^{\prime 2}}, 
\hspace{0.3 cm}  \tilde h = {\chi^2_0 \chi_0^{\prime \prime \prime} \over  \chi_0^{\prime 3}}.
\end{eqnarray}
Here $\Omega_{\chi, D}$ is the energy density at time of curvaton decay,
$\chi_0$ is the curvation field during oscillations just before decay and the
primes here denote derivatives with respect to time. 

Unlike single scalar field inflation, curvaton models can have large self
interactions.    Enqvist et al. pointed out that even if $f_{\rm NL}$ is small,
$g_{\rm NL}$ can be large for significant levels of
self-interactions~\cite{Enqvist:2009ww}.  This places a physical significance
on $g_{\rm NL}$ that can be thought of as parameterizing large
self-interactions.

%
%

\subsection{Brief Summary}

In this section we have discussed the physical significance of each statistic
$f_{\rm NL}$, $g_{\rm NL}$, $\tau_{\rm NL}$ and $A_{\rm NL}$.  In the
bispectrum,  $f_{\rm NL}$ receives contributions from different shaped
triangles in Fourier space related to different underlying physics.  By
analyzing the amount of non-Gaussianity from these different shapes we can
distinguish between models with multiple fields, non-canonical kinetic energy
and non-adiabatic vacuums.  

In addition we stressed the physical significance of local non-Gaussianity in
the trispectrum. The relation $A_{\rm NL} = \tau_{\rm NL}/(6 f_{\rm NL}/5)^2$
is an important constraint of multi-field models.  A general result for single-field 
models is $A_{\rm NL} = 1$.  Lastly, $g_{\rm NL}$ will place important
constraints on the level of self-interactions.

%
%

\section{Power Spectra Estimators For First and Second-Order Non-Gaussianity.}
\label{sec:Theory}

 We would like to find a way to measure the non-Gaussianity of these fields
from something directly observable.  Fortunately, information about the
curvature perturbations are contained within the CMB through the spherical
harmonic coefficients of the temperature anisotropies:
\begin{eqnarray}
a_{lm} &=& 4 \pi (-i)^l \int \frac{d^3{\bf k}}{(2 \pi)^3} \Phi({\bf k})  g_{Tl}(k)Y_{l}^{m*}({\hat{\bf k}}), \\ 
 \theta(\hat{\bf n}) &=& \frac{\delta T}{T}(\hat{\bf n}) = \sum_{l m} a_{lm}Y_{l}^{m*}({\hat{\bf n}})\, ,
\label{eq:alm} 
\end{eqnarray}
where $\Phi({\bf k})$ are the primordial curvature perturbations, $g_{Tl}$ is
the radiation transfer function that gives the angular power spectrum $C_l = (2 / \pi) \int k^2 dk P_{\Phi}(k)g^2_{{\rm T}l}(k)$, $\theta$ is the field of temperature
fluctuations in the CMB and $Y^{l}_m$'s are the spherical harmonics. (In this
equation, the curvature perturbation $\Phi$ is related to $\zeta$ through the
relation $\Phi = (3/5)\zeta$.)

If the curvature perturbations are purely Gaussian, all the statistical information we can say about them is contained in the two point correlation function $\left<\Phi({\bf x_1}) \Phi({\bf x_2})\right> $. The information contained in the two point function is usually extracted in spherical harmonic space, leading to the power spectrum $C_l$, defined by: 
\begin{eqnarray} C_l = \left<a_{lm} a_{lm}\right> = {1 \over (2l+1)}\sum_m
a_{lm} a^*_{lm} \label{eq:powerspec}. 
\end{eqnarray}

However, if the curvature perturbations are slightly non-Gaussian, this two
point function is no longer sufficient to articulate all the information
contained in the field.  With non-Gaussianity, extra information can be
extracted from the three, four and higher n-point correlation
functions~\cite{Komatsu:2009kd}.  

We now discuss estimators that can be used to measure non-Gaussianity at first
and second order corresponding to the third and fourth-order in curvature
perturbations respectively.

%
%

\subsection{Skewness Power Spectrum Estimator for the Bispectrum.}

In order to detect Gaussianity at first order, we must turn to the three point
correlation function of the primordial curvature perturbations $\left<\Phi({\bf
x_1}) \Phi({\bf x_2}) \Phi({\bf x_3})\right> $.  As mentioned above, we can
extract information from the curvature perturbations by analyzing the
$a_{lm}$s of the CMB.  The three point correlation function of the $a_{lm}$s is
called the bispectrum can be decomposed as follows\cite{Komatsu:2002db}:
\begin{eqnarray}
\label{eq.Bispectrum}
\langle a_{lm}a_{l'm'}a_{l''m''} \rangle
= B_{ll'l''} \left ( \begin{array}{ c c c }
     l & l' & l'' \\
     m & m' & m''
  \end{array} \right).
\end{eqnarray}
where
\begin{eqnarray}
\label{eq:bispectra}
B_{ll'l''} \equiv  \sqrt {(2l+1)(2l'+1)(2l'+1)\over 4\pi}\left ( \begin{array}{ c c c }
     l & l' & l'' \\
     0 & 0 & 0
  \end{array} \right)b_{ll'l''}.
\end{eqnarray}

Here the  symbols in parenthesis are called the Wigner-3j symbols and enforce
rotational invariance of the CMB, as well as ensuring the proper triangle
equality holds between $l$, $l'$ and $l''$ namely: $|l_i -l_j| \le l_k \le |
l_i + l_j |$ for any combination of $i$, $j$ and $k$. For more information on
the wigner 3j symbols, the reader is directed to the appendix of
Ref.~\cite{Komatsu:2002db}. 

The quantity $b_{ll'l''}$, known as the reduced bispectrum, encases all the
other information in the bispectrum and for the local model can be computed
analytically as:
\begin{eqnarray}
\label{eq:reducedbispec}
&&\hspace{-0.6cm}b_{l_1l_2l_3} = 2 f_{NL}^{\rm} \int r^2 dr \left [ \alpha_{l_1}(r) \beta_{l_2}(r) \beta_{l_3}(r) + {\rm cyc.perm.} \right ],
\end{eqnarray}
where
\begin{eqnarray}
\label{eq:alphabeta}
\alpha_l(r) &\equiv& {2 \over \pi} \int_0^{\infty} k^2 dk g_{Tl}(k) j_l(kr),  \nonumber\\
\beta_l(r) &\equiv &  {2 \over \pi} \int_0^{\infty} k^2 dk P_{\Phi}(k) g_{Tl}(k) j_l(kr). 
\end{eqnarray}
Here, $P_\Phi(k)\propto k^{n_s-4}$ is the primordial power spectrum of
curvature perturbations,  $g_{Tl}(k)$ is defined above, $j_l(kr)$ are the
spherical Bessel functions and $r$ parameterizes the line of sight. 

\begin{figure}
    \begin{center}
      \includegraphics[scale=0.45]{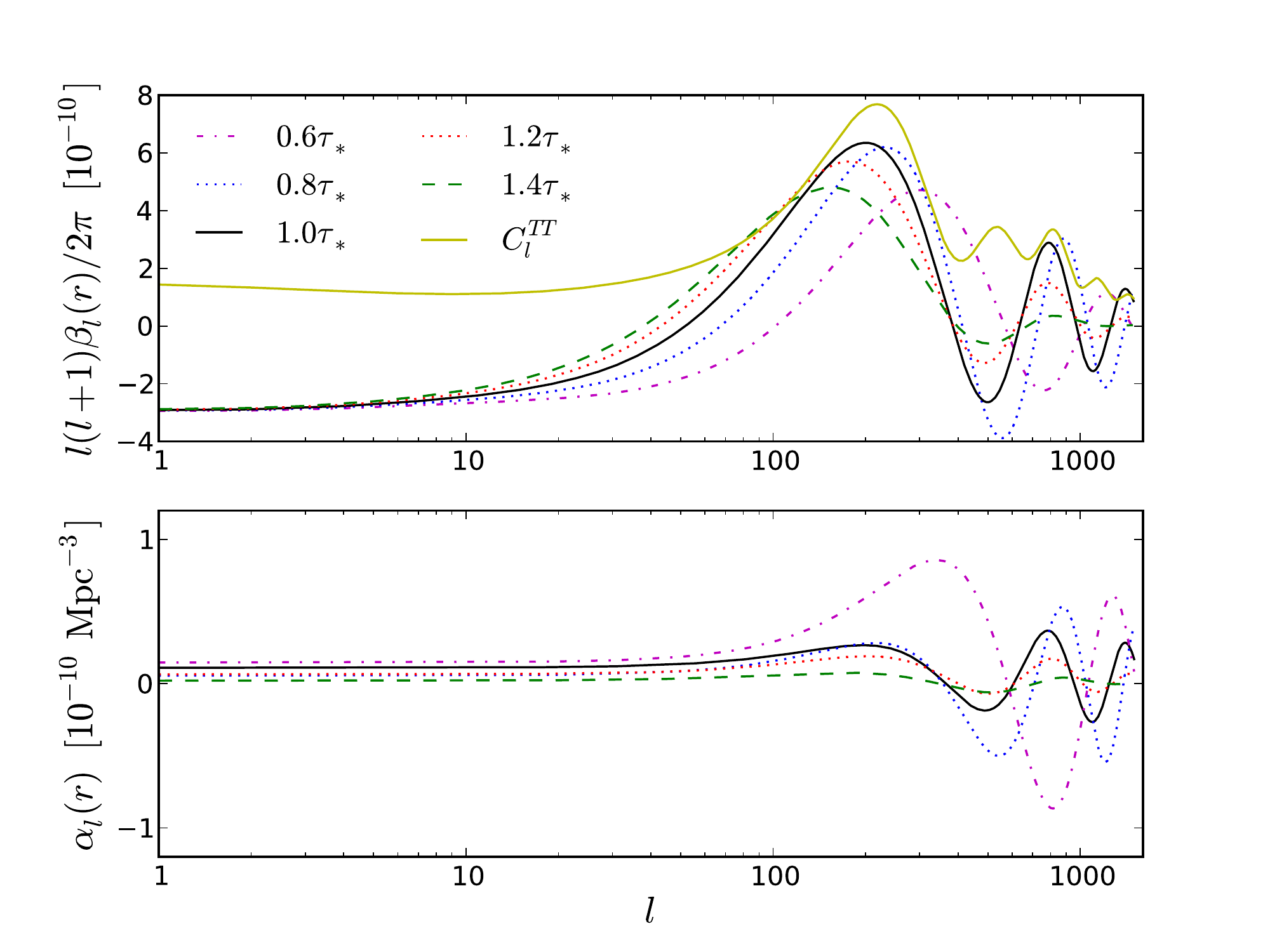} 
   \end{center}
   \vspace{-0.7cm}
   \caption[width=3in]{The top plot compares various $\beta(r)$ for different $\tau_{*}$ and the bottom is the same for $\alpha(r)$.}
   \label{fig:alphabeta}
\end{figure}

Traditionally, the non-Gaussianity parameter $f_{\rm NL}$ is given as a single number where the information from all the triangles configurations are
collapsed into a single number called skewness ($S_3$) defined as
\begin{equation}
S_3  \equiv \int r^2 dr \int d \n  A(r,\n )B^2(r,\n). 
\end{equation}

where $A(r,\n )$ and $B(r,\n)$, are defined below in equations~\ref{eq:Amap}
and~\ref{eq:Bmap}.  Recently, new techniques have been developed to measure
$f_{\rm NL}$ from a power spectrum called the skewness power
spectrum~\cite{Cooray:2001ps,Munshi:2009ik}.  These new estimators based on the
analysis of power spectra are equivalent  to $S_3$ in the limit of homogeneous
noise~\cite{Komatsu:2010fb} but have certain advantages discussed at the bottom
of this subsection.  These advantages include the ability to separate
foregrounds and other secondary non-Gaussian signals and the ability to correct
for the cut-sky without having to compute so-called linear terms.

To extract the skewness power spectrum from data we must begin with temperature
maps optimally weighted for the detection of non-Gaussianity
following~\cite{Komatsu:2003iq}:
\begin{eqnarray}
\label{eq:Amap}
\hspace{-0.4cm}A(r,\n) \hspace{-0.1cm}&\equiv& \hspace{-0.1cm}\sum_{lm} Y_{lm}(\n) A_{lm}(r); ~ A_{lm}(r) \equiv {\alpha_{l}(r) \over {\cal C}_l} b_l a_{lm}, \\
\label{eq:Bmap}
\hspace{-0.4cm}B(r,\n) \hspace{-0.1cm}&\equiv&\hspace{-0.1cm} \sum_{lm} Y_{lm}(\n) B_{lm}(r); ~ B_{lm}(r) \
\equiv {\beta_{l}(r) \over {\cal C}_l} b_l a_{lm}.\end{eqnarray}
Here ${\cal C}_l \equiv C_l b_l^2 + N_l$ where $b_l$ and $N_l$ are the beam
transfer functions and noise power spectrum respectively as described below in
Section~\ref{sec:realexp} and $C_l$ is the usual two point correlation function
defined above in equation~\ref{eq:powerspec}.

From the two above weighted maps we can create two unique two-one power
spectra, each of which contribute to the full $C_l^{(2,1)}$ estimator defined
as:
\begin{eqnarray}
\label{eq:CAB2}
C_l^{A,B^2} &\equiv& \int r^2 dr~ C_l^{A,B^2}(r), \\
\label{eq:CABB}
C_l^{AB,B} &\equiv& \int r^2 dr~ C_l^{AB,B}(r),
\end{eqnarray}
where
\begin{eqnarray}
C_l^{A,B^2}(r) &= &{1 \over 2l + 1} \sum_m {\rm Real}\left \{ A_{lm}(r) B_{lm}^{(2)}(r) \right \}; \\
C_l^{AB,B}(r) &= &{1 \over 2l + 1} \sum_m {\rm Real}\left \{ (AB)_{lm}(r) B_{lm}(r) \right \}.
\end{eqnarray}
It should make sense that the integrals with respect to the line of sight are
needed since the final power spectra must only be an $l$ dependent quantity.

In the above equations, the {\it squared} multipole moments are defined in
relation to the squared optimized temperature maps as:
\begin{eqnarray}
\label{eq:B2lm}
 B(r,\n)^2 &=& \sum_{lm} B_{lm}^{(2)}(r)Y_{l}^{m*}({\hat{\bf n}}); \\
 \label{eq:ABlm}
 A(r,\n) B(r,\n)&=& \sum_{lm} (AB)_{lm}(r)Y_{l}^{m*}({\hat{\bf n}}).
\end{eqnarray}

Combining the two unique contributions from equations~\ref{eq:CAB2}
and~\ref{eq:CABB} gives us our full skewness power spectrum estimator:
\begin{equation}
C_l^{(2,1)} \equiv (C_l^{A,B^2} + 2 C_l^{AB,B}). 
\label{eq:fnlee}
\end{equation}

Once $C_l^{(2,1)}$ has been extracted from data, we can compute the amount of
non-Gaussianity found therein by relating this estimator to its analytical
expression for a model with $f_{\rm NL} = 1$ that turns out to be:
\begin{equation}
C_l^{(2,1)} = {f_{NL} \over (2l+1)} \sum_{l'}\sum_{l''} \left \{ {B_{ll'l''} \hat B_{ll'l''} \over {\cal C}_l {\cal C}_{l'} {\cal C}_{l''}} \right \}. 
\label{eq:fnle}
\end{equation}
Here, ${\cal C}_l$ is the weighted two point power spectrum defined below
equation~\ref{eq:Bmap}, $\hat B_{ll'l''}$ is the full bispectrum and
$B_{ll'l''}$  is the local model with $f_{\rm NL} = 1$ calculated from
equations~\ref{eq:bispectra} and~\ref{eq:reducedbispec}.

Measuring non-Gaussianity using a power spectrum has a few advantages related
to the fact that all information is not squeezed into a single number.  First,
different physics that contribute to the  bispectrum, such as point sources and
secondaries, can be directly accounted for and measured using curve fitting
techniques utilizing each quantities two-one spectrum and fitting all
parameters simultaneously as was done recently in Smidt et. al
2009~\cite{Smidt:2009ir}.  Second, each statistic can be tested for scale
dependance with ease.  This was also done in~\cite{Smidt:2009ir} where it
was found that $f_{\rm NL}$ is consistant with zero for all $l$.   Third,
effects due to the cut sky can be removed easily without needing to calculate
linear terms needed with $S_3$. We discuss this later issue in
Section~\ref{sec:masksky}.  Lastly, for the trispectrum analysis discussed
below, both second order statistics $\tau_{\rm NL}$ and $g_{\rm NL}$ can be
calculated simultaneously using the two kurtosis spectra.      

%
%

\subsection{Kurtosis Power Spectrum Estimators for the Trispectrum}

In order to extract  non-Gaussianity at second order we must consider the
trispectrum or four point function of temperature anisotropies which
conveniently breaks into a Gaussian and non-Gaussian or connected
piece~\cite{Hu:2001fa}:
\begin{eqnarray}
\left<a_{l_1m_1} a_{l_2m_2} a_{l_3m_3} a_{l_4m_4}\right> &=& \\
\left<a_{l_1m_1} a_{l_2m_2} a_{l_3m_3} a_{l_4m_4}\right>_G\,   
&+& \left<a_{l_1m_1} a_{l_2m_2} a_{l_3m_3} a_{l_4m_4}\right>_c.\nonumber
\label{eq:tripieces}\end{eqnarray}
where the connected and unconnected part of the trispectrum can be expanded as:
\begin{eqnarray}\label{eq:triconnected}
\left <a_{l_1m_1} a_{l_2m_2} a_{l_3m_3} a_{l_4m_4}\right>_G =&& \\ 
\sum_{L M} (-1)^{M} G_{l_1 l_2}^{l_3 l_4} (L)  \wjmL &&  \hspace{-0.5cm}  \wjmmL, \nonumber \\
\nonumber \\
\left <a_{l_1m_1} a_{l_2m_2} a_{l_3m_3} a_{l_4m_4}\right>_c =&& \\ 
\sum_{L M} (-1)^{M} T_{l_1 l_2}^{l_3 l_4} (L)  \wjmL &&  \hspace{-0.5cm}  \wjmmL, \nonumber 
\end{eqnarray}
where we can solve for $G_{l_1 l_2}^{l_3 l_4} (L)$ and $T_{l_1 l_2}^{l_3 l_4}
(L)$ analytically as:
\begin{eqnarray}
\label{eq:reducedtri}
T^{l_1l_2}_{l_3l_4}(L) = (5 / 3)^2 \tau_{\rm NL} h_{l_1l_2L}h_{l_3l_4L} \times &&  \\
 \int r_1^2dr_1 r_2^2 dr_2 F_L(r_1,r_2) \alpha_{l_1}(r_1)&& \hspace{-0.4cm}\beta_{l_2}(r_1)\alpha_{l_3}(r_2)\beta_{l_4}(r_2) \nonumber \\
+ g_{\rm NL} h_{l_1l_2L}h_{l_3l_4L} \times && \nonumber \\
\int r^2 dr \beta_{l_2}(r) \beta_{l_4}(r) [ \mu_{l_1}(r)\beta_{l_3}(r) && \hspace{-0.4cm}+ \mu_{l_3}(r)\beta_{l_1}(r)], \nonumber
\end{eqnarray}
\begin{eqnarray}
\label{eq:gausspiece}
G_{l_1 l_2}^{l_3 l_4} (L) = (-1)^{l_1+l_3} && \hspace{-0.7cm}  \sqrt{(2 l_1+1)(2 l_2+2)} 
C_{l_1} C_{l_3} \delta_{L 0} \delta_{l_1 l_2}\delta_{l_3 l_4} \nonumber \\
(2 L +1) C_{l_1} C_{l_2}  && \hspace{-0.5cm} \left[(-1)^{l_2 + l_3 +L}\delta_{l_1 l_3} \delta_{l_2 l_4} + \delta_{l_1 l_4}\delta{l_2 l_3}\right],
\end{eqnarray}
with $\tau_{\rm NL}$ and $g_{\rm NL}$ being parameters of second order
primordial non-Gaussianity (see discussion in next subsection for more
information).  Written in this form, $T^{l_1l_2}_{l_3l_4}(L)$ above is called the reduced trispectrum  and contains all the physical information about non-Gaussian sources~\cite{Kogo:2006kh}.  The full trispectrum, in general, contains additional terms based on permutations of $l_i$. We approximate the full trispectrum with the reduced trispectrum since
we will be optimizing the estimator with weights to measure a single term of the full trispectrum. There are additional cross terms in our analysis that we then ignore.
The approximation we implement here is already  costly computationally and the lack of including extra cross terms associated with permutation, at most, causes our error 
bars on the non-Gaussian parameters to be overestimated.  Furthermore, as we measure non-Gaussian parameters using the reduced trispectrum, we can
direcly compare our results with the previous predictions that also utilized the same approximation~\cite{Kogo:2006kh,Okamoto:2002ik}. 

In above, the quantity  $h_{l_1l_2l_3}$ is defined such that
\begin{equation}
h_{l_1l_2l_3} = \sqrt{(2l_1+1)(2l_2+1)(2l_3+1)\over 4 \pi}
\left ( \begin{array}{ c c c }
     l_1 & l_2 & l_3 \\
     0 & 0 & 0
  \end{array} \right), \\
\end{equation}
 and\vspace{-0.5cm}
\begin{eqnarray}
\label{eq:FL}
 F_L(r_1,r_2)&\equiv&\frac{2}{\pi}\int k^2dk  P_\Phi(k)j_L(kr_1)j_L(kr_2).
\end{eqnarray}
Here, $P_\Phi(k)\propto k^{n_s-4}$ is the primordial power spectrum of
curvature perturbations, the $\alpha(r)$, $\beta(r)$ and $g_{Tl}(k)$ are
defined above and $j_l(kr)$ are the spherical bessel functions and $r$
parameterizes the line of sight.  

As with the bispectrum, we would like to figure out how to calculate power
spectra that can be related to analytical expressions proportional to
$\tau_{\rm NL}$ and $g_{\rm NL}$.  To do this we begin with the same weighted
maps defined in equations~\ref{eq:Amap} and~\ref{eq:Bmap} which leads to the
spectra:
\begin{eqnarray}
\label{eq:K31}{\cal K}_l^{(3,1)} &=& (5/3)^2\tau_{\rm NL} {\cal J}_l ^{ABA,B} + 2g_{\rm NL}  {\cal L}_l^{AB^2,B}, \\
\label{eq:K22}{\cal K}_l^{(2,2)} &=& (5/3)^2\tau_{\rm NL}  {\cal J}_l^{AB,AB} + 2g_{\rm NL}  {\cal L}_l^{AB,B^2}, 
\end{eqnarray}
where the unique two-two and three-one power spectra are:
\begin{eqnarray}
\label{eq:JA2BB} {\cal J}_l ^{ABA,B} &=&  \int r_1^2 dr_1 \int r_2^2 dr_2  {\cal J}_l
^{ABA,B}(r_1,r_2); \\
\label{eq:LAB2B}
{\cal L}_l^{AB^2,B} &=& \int r^2 dr {\cal L}_l^{AB^2,B}(r); \\
\label{eq:JABAB}{\cal J}_l^{AB,AB}  &=&  \int r_1^2 dr_1 \int r_2^2 dr_2 {\cal J}_l^{AB,AB}(r_1,r_2); \\
\label{eq:LABB2}
 {\cal L}_l^{AB,B^2} &=&  \int r^2 dr {\cal L}_l^{AB,B^2}(r). 
\end{eqnarray}
Here ${\cal J}_l^{ABA,B}(r_1,r_2)$, ${\cal L}_l^{AB^2,B}(r)$, ${\cal J}_l^{AB,A
B}(r_1,r_2)$, and ${\cal L}_l^{AB,B^2}(r)$ are the angular power spectra of
their respective maps. For example ${\cal L}_l^{AB^2,B}(r)$ is defined as:
\begin{equation}
\label{eq:Lpowspec}
{\cal L}_l^{AB^2,B}(r) = {1 \over 2l+1}\sum_m (AB^2)_{lm} B^*_{lm}
\end{equation}
where $(AB^2)_{lm}$ and$ B^*_{lm}$ are defined analogously with
equations~\ref{eq:B2lm} and~\ref{eq:ABlm}.

Once the kurtosis estimators have been extracted from temperature data, we can
fit the two unknowns $\tau_{\rm NL}$ and $g_{\rm NL}$ from the two estimators
simultaneously by comparing them to their analytical expressions with
$\tau_{\rm NL} = g_{\rm NL} = 1$ that turn out to be~\cite{Munshi:2009wy}:
\begin{eqnarray}
 \label{eq:k22e} {\myK}_l^{(2,2)} &=&  {1 \over (2l+1)} \sum_{l_i} {1 \over (2l+1)}
 { T_{l_1l_2}^{l_3l_4}(l) \hat T^{l_1l_2}_{l_3l_4}(l) \over {\cal C}_{l_1} {\cal C}_{l_2} {\cal C}_{l_3
} {\cal C}_{l_4}}; \\ \label{eq:k31e}
 {\myK}_l^{(3,1)} &=&   {1 \over (2l+1)}  \sum_{l_iL}  {1 \over (2L+1)}  { T^{l_
1l_2}_{l_3l}(L) \hat T^{l_1l_2}_{l_3l}(L) \over {\cal C}_{l_1} {\cal C}_{l_2} {\cal C}_{l_3} {\cal C}_{l}}.\end{eqnarray}
where $\hat T_{l_1l_2}^{l_3l_4}(l)$ is the full bispectrum and
$T_{l_1l_2}^{l_3l_4}(l)$  is the local model with $\tau_{\rm NL} = g_{\rm NL} =
1$ calculated from equation~\ref{eq:reducedtri}. 

%
%

\section{Fisher bounds}
\label{sec:fishertheory}

%
%

\subsection{The Ideal Experiment}

In order to determine the optimal error bars for these estimators we must
properly calculate their signal-to-noise ratios.  For the bispectrum,  the
signal-to-noise ratio takes on the simple form 
\begin{eqnarray}
 \label{eq:Fc21} 
 \left({S \over N}\right)^2_{(2,1)} &=&  \sum_l (2l+1) C_l^{(2,1)},
 \end{eqnarray}
where $C_l^{(2,1)}$ is defined above in eq~\ref{eq:fnle}.

For the trispectrum we must calculate the signal-to-noise for both
${\myK}_l^{(2,2)}$ and $ {\myK}_l^{(3,1)}$.  In a best case scenario, the two
estimators above are not correlated.  In this case the signal-to-noise for each
estimator is:
\begin{eqnarray}
 \label{eq:Fk22e} 
 \left({S \over N}\right)^2_{(2,2)} &=&  \sum_l (2l+1) {\myK}_l^{(2,2)}; \\ 
 \label{eq:Fk31e}
 \left({S \over N}\right)^2_{(3,1)} &=&  \sum_l (2l+1) {\myK}_l^{(3,1)}.
\end{eqnarray}
Given the positive definite nature of $\left(S / N\right)^2$, the
signal-to-noise increases as one computes to higher $l$ values.  In fact, for
the trispectrum it has been shown that $\left(S / N\right)^2 \sim l_{{\rm
max}}^4$ where $l_{\rm max}$ represents the maximum $l$ used in the
analysis~\cite{Kogo:2006kh}.

In addition to the estimators themselves being correlated, contributions to the
terms proportional to $\tau_{\rm NL}$ and $g_{\rm NL}$ come from different
quadratic contributions in Fourier space.  This further allows us to calculate
the signal-to-noise for each of these terms in each estimator by setting the
other to zero.  For example, we can determine the optimal signal-to-noise for
the $\tau_{\rm NL}$ term from say the  ${\myK}_l^{(2,2)}$ estimator by setting
$g_{\rm NL} = 0$ embedded in equation~\ref{eq:Fk22e}.

Once the signal-to-noise is known, we immediately have a bound on the optimal
error bars for our estimators through the inverse square root.  For example, if
we wanted to know the optimal $1\sigma$ error bar that can be placed on
$\tau_{\rm NL}$ from the  ${\myK}_l^{(2,2)}$ estimator, we can compute the
Fisher bound as
\begin{equation}
\label{eq:fisher}
\sigma(\tau_{\rm NL}) = {1 \over \sqrt{\left({S \over N}\right)_{(2,2)}^2|_{\tau_{\rm NL}}}} \, ,
\end{equation} 
with the restriction on $\left(S / N\right)^2_{(2,2)}$ to $\tau_{\rm NL}$ by
setting  $g_{\rm NL} = 0$ in this calculation.

%
%

\subsection{The Realistic Experiment.}
\label{sec:realexp}

In the above equations we assumed a ``perfect'' experiment with no noise or
beam with a full sky.  We now must take in to account that real world
experiments have an inherent noise associated with the detector and a beam to
characterize its angular resolution.  Both the noise and the beam reduce the
signal-to-noise.  Furthermore, the mask yields a cut sky that must be dealt
with properly.  

The noise is often reasonably approximated assuming a homogeneous spectrum
calculating $N_l$ from the following relation:
\begin{eqnarray}
N_l &=& \sigma_{\rm pix}^2  \Omega_{\rm pix}, 
\end{eqnarray}
where $\sigma_{\rm pix}$ is the rms noise per pixel and $\Omega_{\rm pix}$ is
the solid angle per pixel.  

For the noise calculation taking into the inhomogeneous coverage of real world
experiments and a cut sky $N_l$ is to be calculated by:
\begin{equation}
N_l = \Omega_{\rm pix} \int {d^2{\bf \hat n} \over 4 \pi f_{\rm sky}} {\sigma_{\rm pix}^2 M({\bf \hat n}) \over N_{\rm obs}({\bf \hat n})},
\end{equation}
where $f_{\rm sky}$ is the fraction of sky observed and $N_{\rm obs}$ is the
number of observations per pixel~\cite{Komatsu:2010fb}.

In addition to noise, realistic detectors have limits to their resolving power.
The resolution limits of the instrument, encoded in the parameter $\theta_{\rm
FWHM}$ which represents the full-width-half-max of the resolving power.  We can
map this information into harmonic space in the beam transfer function $b_l$ 
\begin{eqnarray}
b_l &=& \exp{\left(-l^2 \sigma_{\rm beam}^2\right)}, \\
\sigma_{\rm beam} &=& {\theta_{\rm FWHM} \over \sqrt{(8 \ln(2))}}.
\end{eqnarray}

\begin{figure}
    \begin{center}
      \includegraphics[scale=0.45]{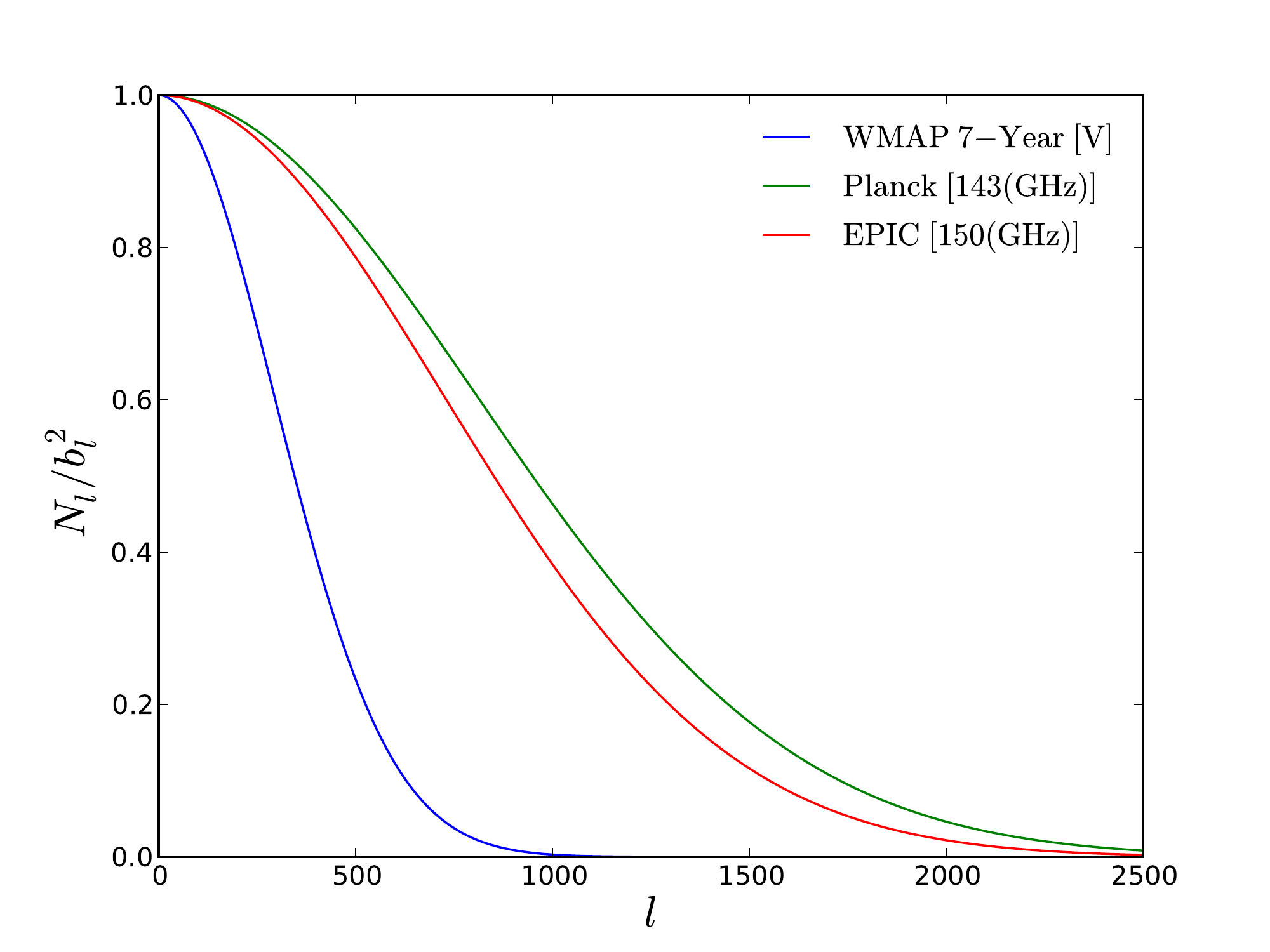} 
   \end{center}
   \vspace{-0.7cm}
   \caption[width=3in]{Beam transfer functions.  The frequency band used for each experiment is in brackets.}
   \label{fig:bl}
\end{figure}

Beam transfer functions for the WMAP, Planck and EPIC experiments are plotted
in Figure~\ref{fig:bl}.   As one would expect, a larger $\theta_{\rm FWHM}$
results in the suppression of information on larger scales.

\begin{table}[htbp]  \centering  \begin{tabular}{@{} |c|c|c|c|c| @{}}    \hline    
Mission & $\theta_{\rm FWHM}$ &  $\sigma_{\rm pix}$ & $\Omega_{\rm pix}$ & Frequency \\    
 \hline    
Planck & 7.1' & $2.2 \times 10^{-6}$ & 0.0349 & 143 (GHz)\\     
EPIC & 5.0' & $8 \times 10^{-9}$ &  0.002 & 150 (GHz) \\     \hline  
\end{tabular}  \caption{Parameters used to calculate the simulated noise and beam transfer functions for the Planck and EPIC experiment~\cite{:2006uk, Baumann:2008aq}.  
We obtained WMAP noise and beam function  from publicly available data.
}  \label{tab:noise}\end{table}
\begin{figure}
    \begin{center}
      \includegraphics[scale=0.45]{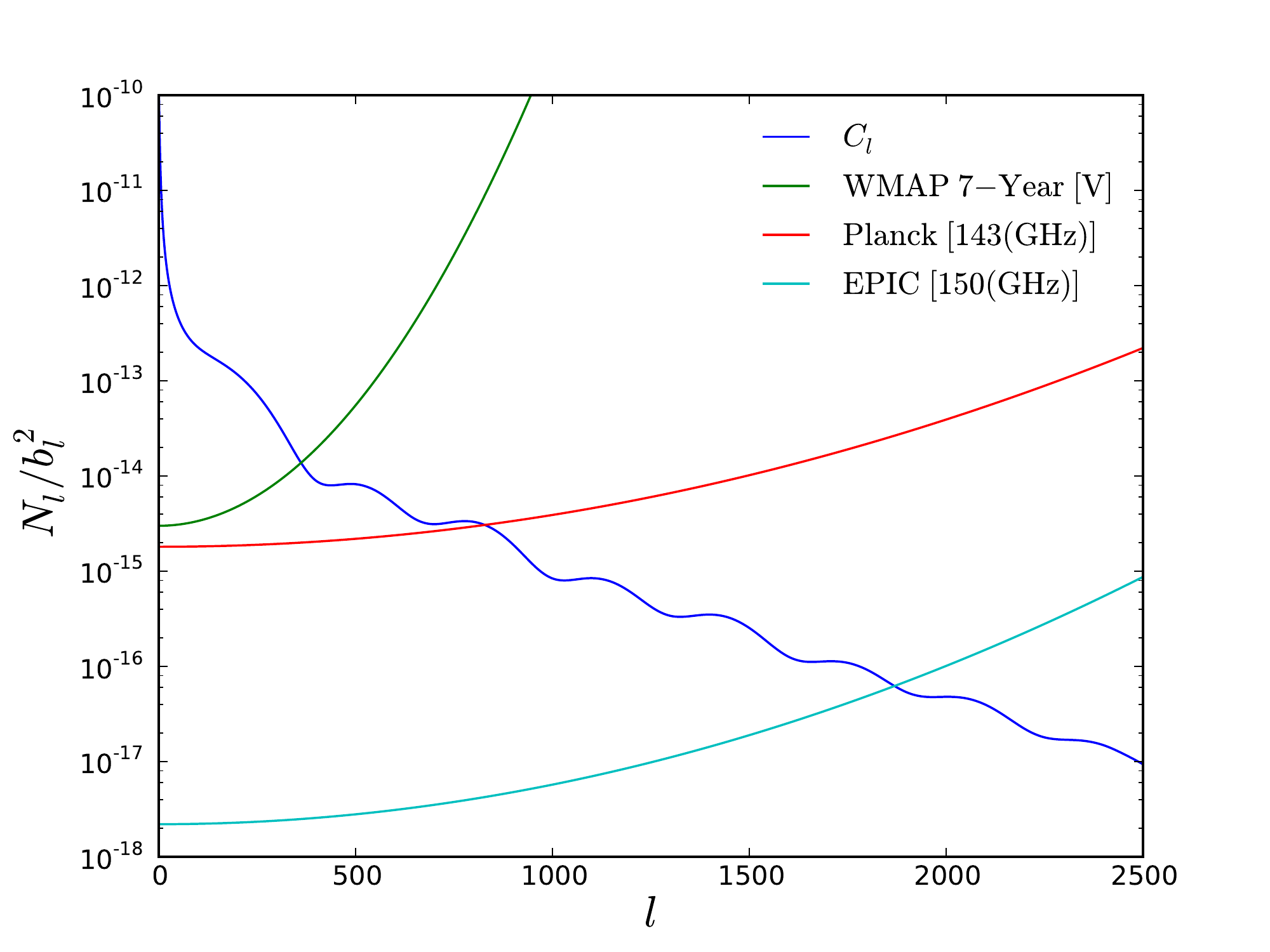} 
   \end{center}
   \vspace{-0.7cm}
   \caption[width=3in]{Noise and $b_l$ relation, $N_l/b_l^2$, for each experiment plotted against $C_l$ taken from WMAP 7-Year best fit parameters.  The frequency band used for each experiment is in brackets.}
   \label{fig:nlbl}
\end{figure}
Working in spherical harmonic space, it is easy to correct our estimators
${\myK}_l^{(2,2)}$ and ${\myK}_l^{(3,1)}$ for the effects due to noise and the
beam.  All that must be done is to preform the transformation:
\begin{equation}
C_L \rightarrow C_l + {N_l \over b_l^2}.
\end{equation}
in the denominator of Eq.~\ref{eq:k22e} and~\ref{eq:k22e}.  As should be
intuitively expected, a large amount of noise, or poor resolution will result
in a smaller signal-to-noise.  Therefore, how much the signal-to-noise is
effected is related to the relationship between $C_l$ and $N_l/b_l^2$.  For
$N_l/b_l^2 >> C_l$, the signal is greatly diminished.  The relation between
$C_l$ and $N_l/b_l^2$ for the WMAP, Planck and EPIC experiments is plotted in
Figure~\ref{fig:nlbl}

%
%

\subsection{Mask And Cut Sky}
\label{sec:masksky}

To remove cut sky effects using the traditional $S_3$ estimator, many linear
terms must be computed that must be subtracted
off~\cite{Creminelli:2005hu,Yadav:2007ny,Komatsu:2008hk}.  Furthermore, the
number of terms that must be computed grows for higher n-correlation functions.
The difficulty arises because the cut sky effects are compressed into a single
number, making it difficult to subtract out.

One advantage of probing primordial non-Gaussianity with skewness power spectra
is that we can use techniques pioneered by Hivon et al. to remove mask effects
from the spectra~\cite{Hivon:2001jp}.  This technique is relatively simple and
works identically for correlation functions of arbitrary order.  

When one uses realistic data, a mask $W({\bf n})$ must be applied to an all sky
map $M({\bf n})$ to get rid of unwanted sources such as the galactic plane.
This mask therefore affects the $a_{lm}$s derived from the all sky $A(r,\n)$
and $B(r,\n)$ defined in equations~\ref{eq:Amap} and~\ref{eq:Bmap} used in the
bispectrum and trispectrum analysis producing cut sky $\tilde{a}_{l m}$s:

\begin{eqnarray}
        \tilde{a}_{l m} &=& \int d{\n} M(\n) W({\bf n}) Y^{m*}_{l}(\n),\\
                       &=& \sum_{l' m'} a_{l' m'} \int d{\n}  Y^{m'}_{l'}({\n}) W(\n) Y^{m*}_{l }({\n}), \\
                       &=& \sum_{l' m'} a_{l' m'} K_{l ml'm'}[W], 
\end{eqnarray}

Here $a_{l'm'}$ is for the full sky,  $M(\n)$ represents an arbitrary full sky
map and $K_{\ell ml'm'}[W]$ now contains all the cut sky information.  

\begin{figure}
    \begin{center}      \includegraphics[scale=0.45]{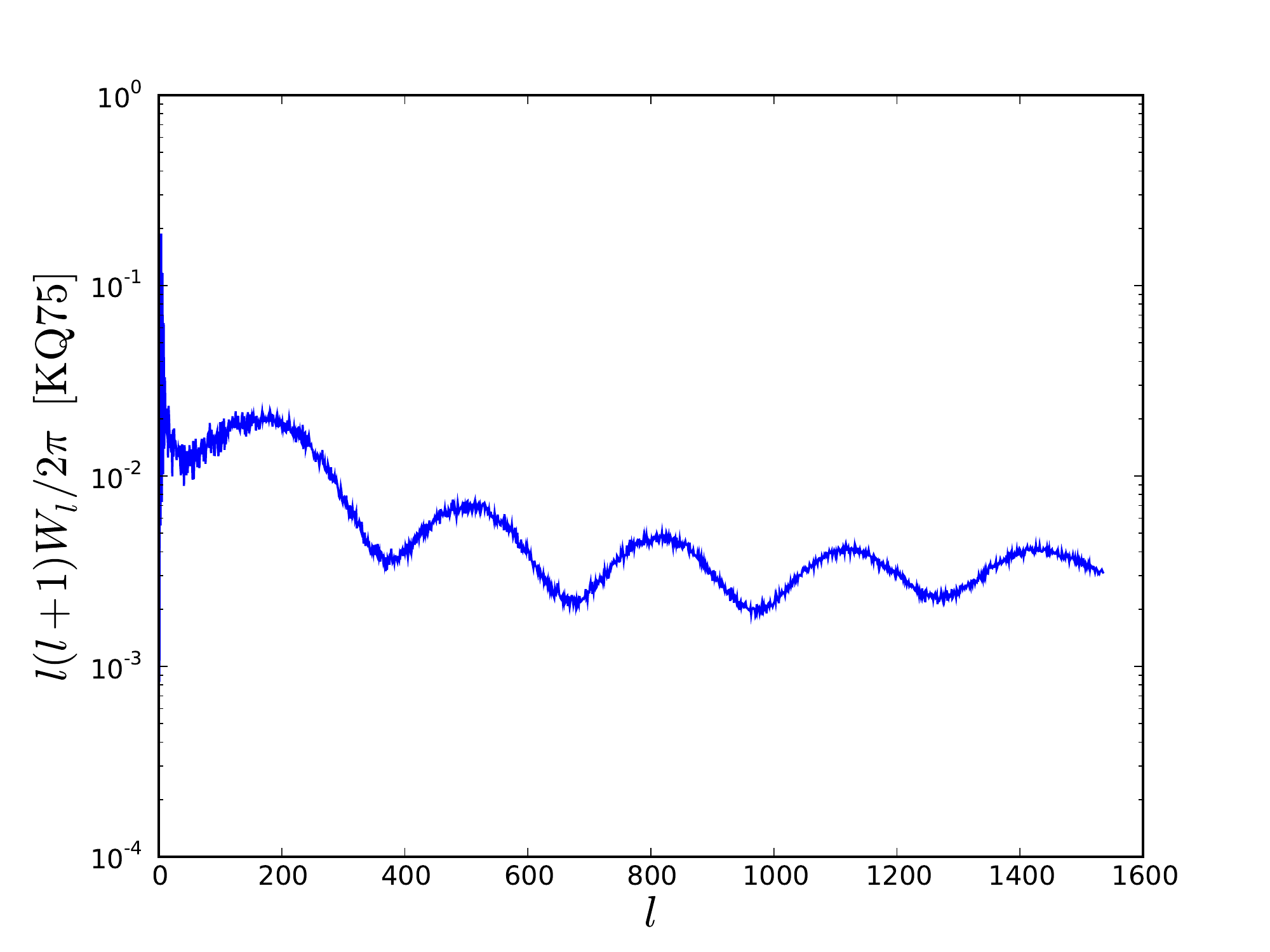}
   \end{center}
   \vspace{-0.7cm}
   \caption[width=3in]{The power spectrum $W_l$ of the KQ75 mask.}
   \label{fig:mask_cl}
\end{figure}

\begin{figure}    \begin{center}
      \includegraphics[scale=0.45]{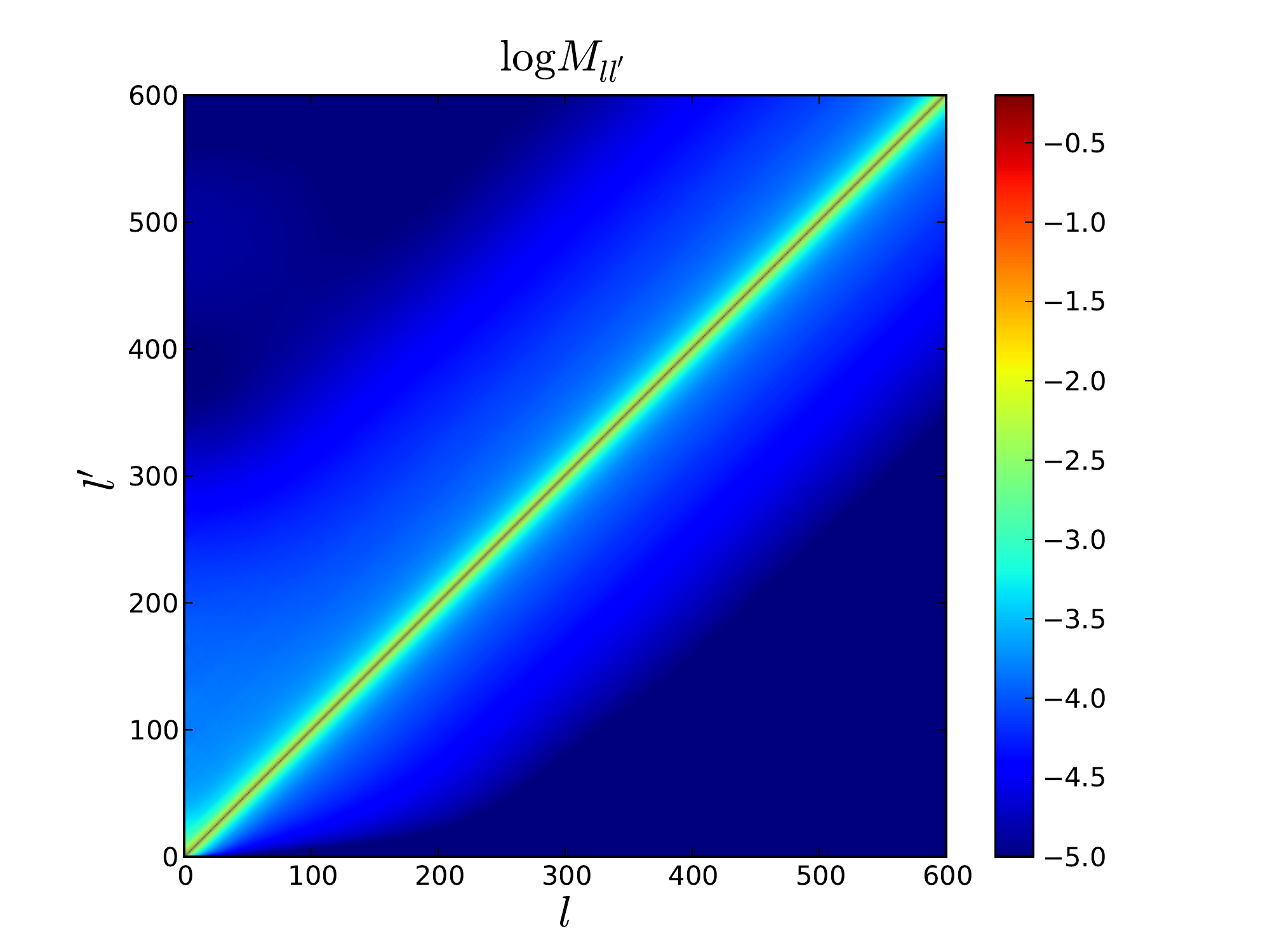}
   \end{center}   \vspace{-0.7cm}   \caption[width=3in]{The matrix $M_{l l'}$ used for correcting the cut sky taken from the KQ75 mask.}   \label{fig:Mllp}\end{figure}

Hivon et al. showed that a power spectrum based on such masked data can be
corrected by:
\begin{equation}
\tilde{C}_l = \sum_{l'} M_{l l'} C_{l'},
\end{equation}
where $M_{l l'}$ is a matrix defined by
\begin{equation}
        M_{l l'} = \frac{2l'+1}{4\pi}\sum_{l''}        (2l''+1) {W}_{l''} \wjjj{l}{l'}{l''}{0}{0}{0}^2.
        \label{eq:kernel_final}
\end{equation}
Here ${W}_{l}$ is the power spectrum of the mask $W({\bf n})$.  The power
spectrum for the KQ75 mask is plotted in Figure~\ref{fig:mask_cl} and the
corresponding $M_{l l'}$ is plotted in Figure~\ref{fig:Mllp}.


\begin{figure}
    \begin{center}
      \includegraphics[scale=0.45]{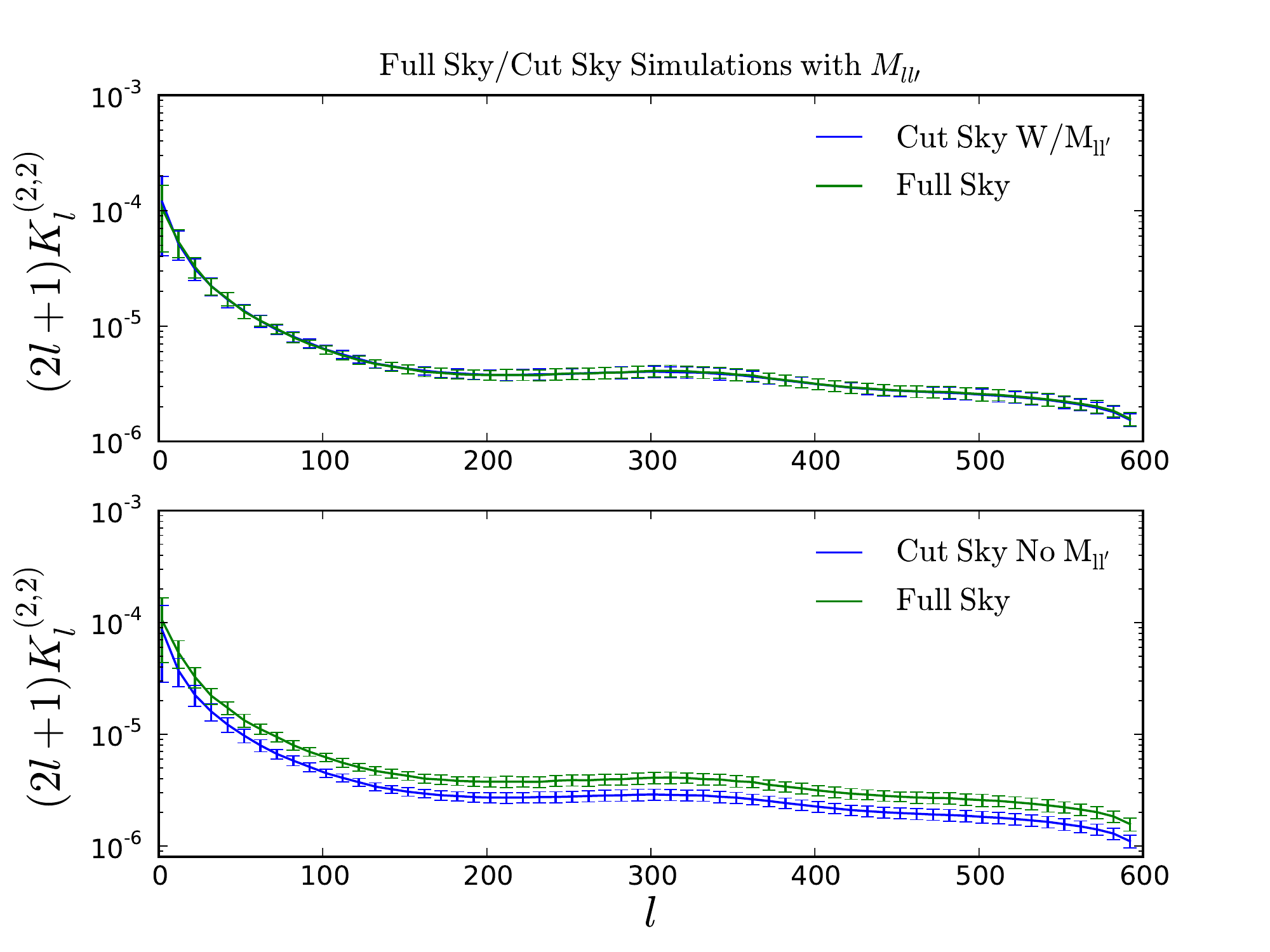}
   \end{center}
   \vspace{-0.7cm}
   \caption[width=3in]{On the bottom we see the results of 250 full sky gaussian simulations of ${\cal K}_l^{(2,2)}$ with the cut sky results without correcting with $M_{l l'}$.  On the top we see the same except the cut sky has received the proper correction.}
   \label{fig:compare_sky}
\end{figure}

Furthermore, it has been shown that any power spectra of rank $C_l^{(p,q)}$ for
any $p$ and $q$ can be corrected with the same method using $M_{l
l'}$~\cite{Munshi:2009wy}.  Thus, we can correct the skewness and kurtosis
power spectrum estimators for the bispectrum (rank $p$ = 2, $q$ = 1) and the
trispectrum (rank $p$ = 2, $q$ = 2 and rank $p$ = 3, $q$ = 1) using this same
technique.  For example, a plot showing the effectiveness of this correction on
the $K_l^{(2,2)}$ estimator is seen in
figure~\ref{fig:compare_sky}.

This correction technique is unique to the power spectrum approach to detect
non-Gaussianity because not all $l$ dependent effects of the mask have
collapsed into a single number.  Therefore, the ability to correct for the mask
in this approach is much easier and more efficient than calculating linear
terms needed to correct for masking effects in for the traditional skewness
statistic $S_3$. 

%
%

\section{Fisher Analysis and Results}
\label{sec:fisherresults}

We now calculate the signal-to-noise for each of our estimators in order to
give reasonable expectations for non-Gaussianity detection from upcoming
experiments using skewness and kurtosis power spectra.  These constraints
assume only temperature data from one frequency band per experiment is used.  
For the WMAP 7-Year analysis we use the V frequency band and for Planck and
EPIC we use the 143 and 150 GHz frequency bands respectively.  The noise and
beam for the WMAP 7-Year V band was taken from the WMAP team and those 
for Planck and EPIC were computed using the values in Table~\ref{tab:noise} as
described in Section~\ref{sec:realexp} .

It should be noted that combining different frequency bands and adding
polarization can further reduce the expected error.   For example. with the
recent WMAP 7-year findings error bars on $f_{\rm NL}$ from one frequency band,
V or W,  is $\pm 24$ but the full temperature analysis combining V+W bands
gives a reduced error bar of $\pm 21$.  (About a 12.5\% improvement over one
temperature frequency band alone.)

For each of these calculations $C_l$, $\alpha(r)$, $\beta(r)$ and $F_L(r_1,r_2)$
were calculated from Eq.~\ref{eq:alphabeta} and ~\ref{eq:FL} using a modified
version of CAMB based on the WMAP 7-Year best fit cosmological  parameter
values.  The quantities $C_l$, $\alpha(r)$ and $\beta(r)$ are plotted in
Figure~\ref{fig:alphabeta}.

For the bispectrum we can form one skewness power spectrum estimator
$C_l^{(2,1)}$ which places bounds on the first order non-Gaussian parameter
$f_{\rm NL}$.  To calculate the signal-to-noise, we compute Eq.~\ref{eq:Fc21}
from Eq.~\ref{eq:fnle} summing all $l$ up to some $l_{\rm max}$ between $2 < l
< 1000$.  After calculating this signal-to-noise we calculate the $l_{\rm max}$
dependent error bars from the Fisher matrix in eq.~\ref{eq:fisher}.  

\begin{figure}
    \begin{center}
      \includegraphics[scale=0.45]{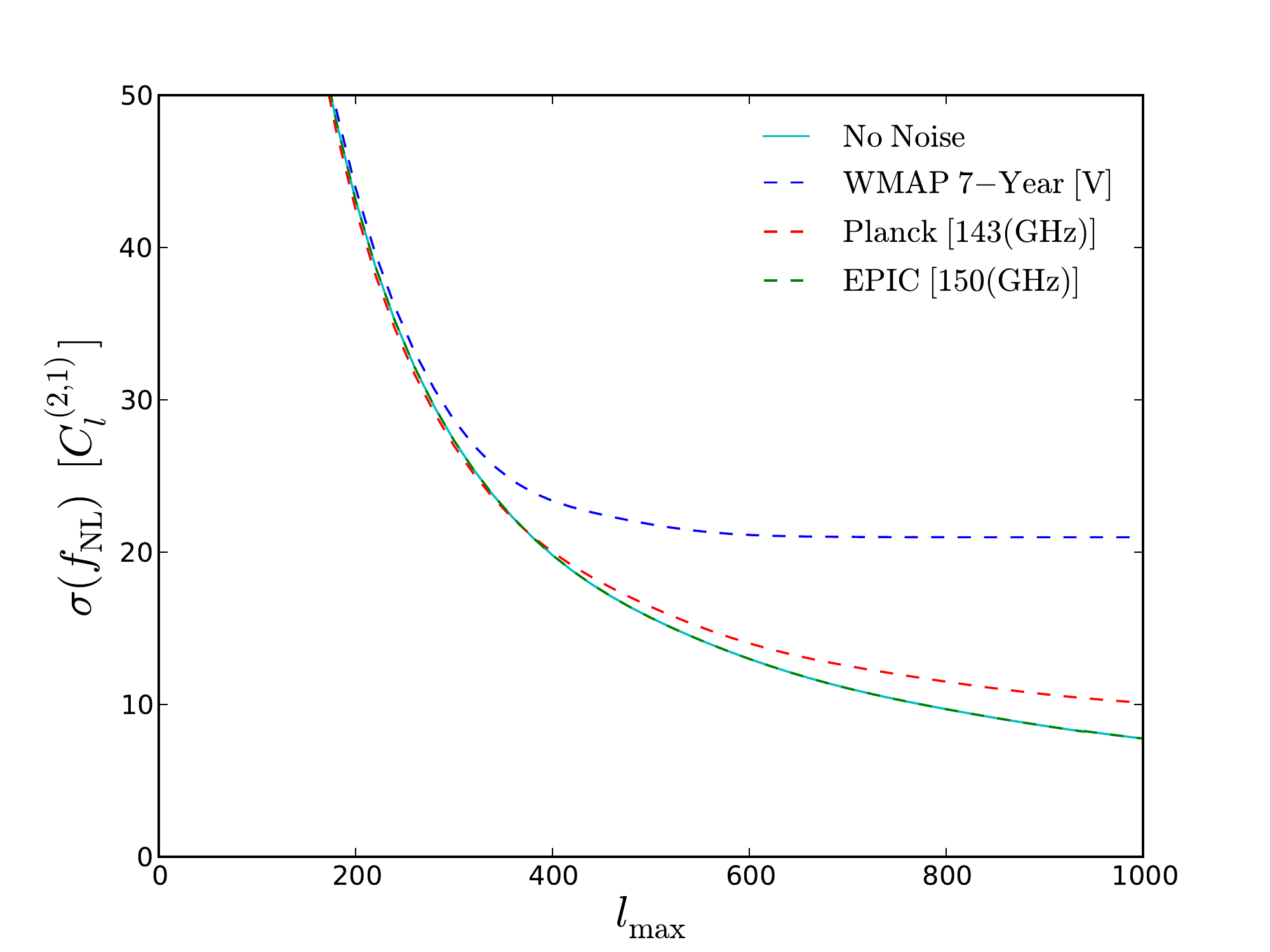} 
   \end{center}
   \vspace{-0.7cm}
   \caption[width=3in]{Fisher bounds on $f_{\rm NL}$ for the ${\cal C}_l^{(2,1)}$ estimator of the bispectrum.  This is calculated from a model with $f_{\rm NL} = 1$. The frequency band used for each experiment is in brackets.}
   \label{fig:c21}
\end{figure}

The results of this calculation are seen in Fig.\ref{fig:c21} and shown in
Table~\ref{tab:sigmabars}.  This calculation is done for the case of no noise
nor beam, as well as with the noise and beam for the experiments WMAP 7-Year,
Planck and EPIC.  As, expected, the error bars drop for higher $l_{\rm max}$
until one reaches the limits of detection for each experiment.  For the case
with no noise, the error bars fall off as $\sim 1/\sqrt{f_{\rm NL} l^2}$. 

For the trispectrum we can form two skewness power spectrum estimators, ${\cal
K}_l^{(2,2)}$ and ${\cal K}_l^{(3,1)}$. For primordial non-Gaussianity
detection that together place bounds on the second order non-Gaussian
parameters $\tau_{\rm NL}$ and $g_{\rm NL}$.  The first of these, ${\cal
K}_l^{(2,2)}$, is computed from eq.~\ref{eq:k22e}.  After this calculation, the
signal-to-noise is computed from eq.~\ref{eq:Fk22e} summing all $l$ up to some
$l_{\rm max}$ between $2 < l < 1000$ for all $l$ except the {\it diagonal} one.
(The diagonal being the $l$ in parenthesis of $T^{l_1l_2}_{l_3l_4}(l)$.)  It
was confirmed, as was previously reported~\cite{Kogo:2006kh}, that nearly all
the signal-to-noise can be calculated only summing up the $l$ in the diagonal
of the trispectrum up to $l = 10$, saving a tremendous amount of computational
time.  In this analysis, however, we summed up the diagonal in both trispectrum
estimators to $l = 20$ so as to be more conservative.   The error bars on
$\tau_{\rm NL}$ and $g_{\rm NL}$ from this estimator are then computed from
equation eq.~\ref{eq:fisher}.
\begin{figure}
    \begin{center}
      {\includegraphics[scale=0.45]{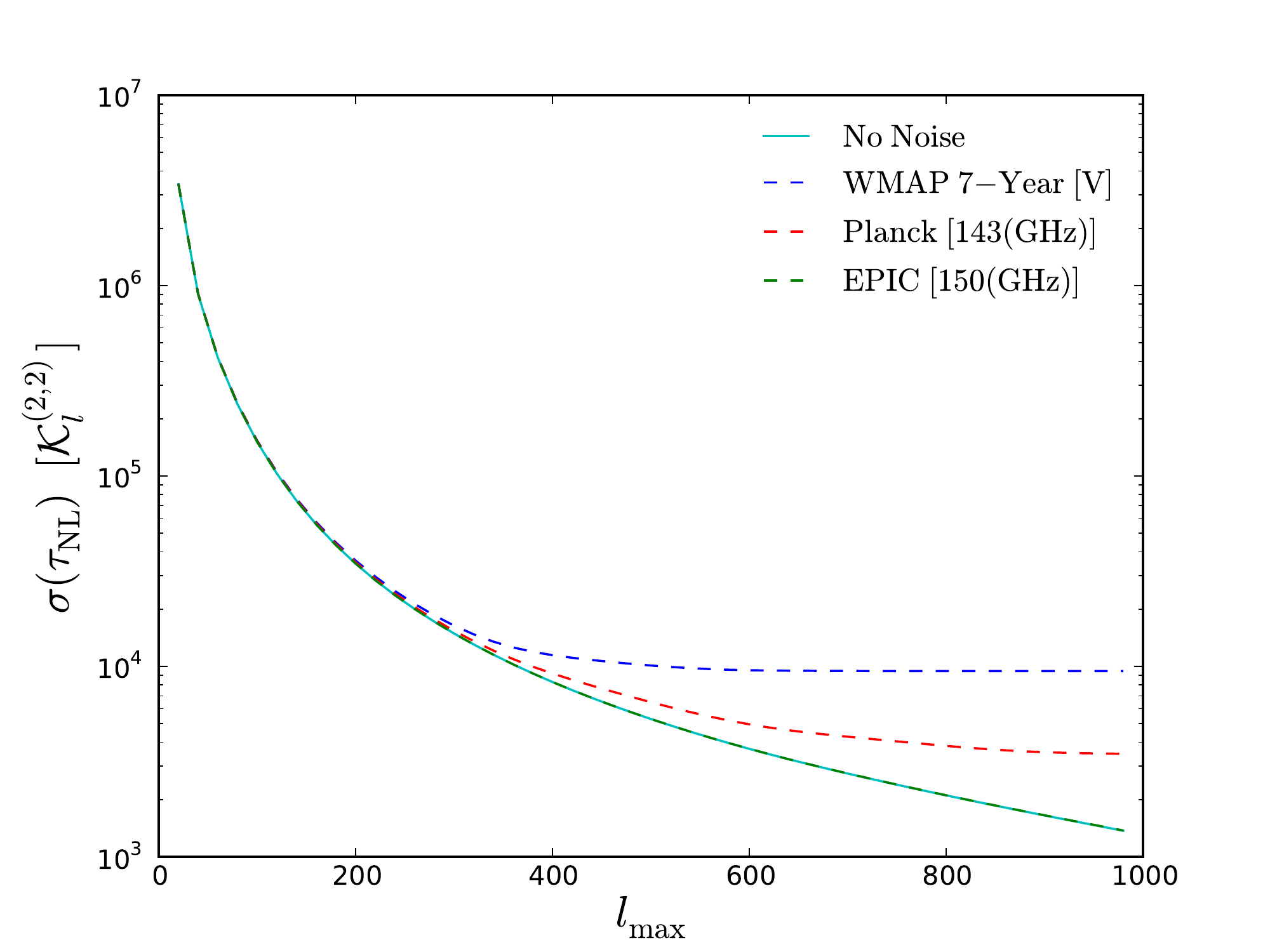} 
      \includegraphics[scale=0.45]{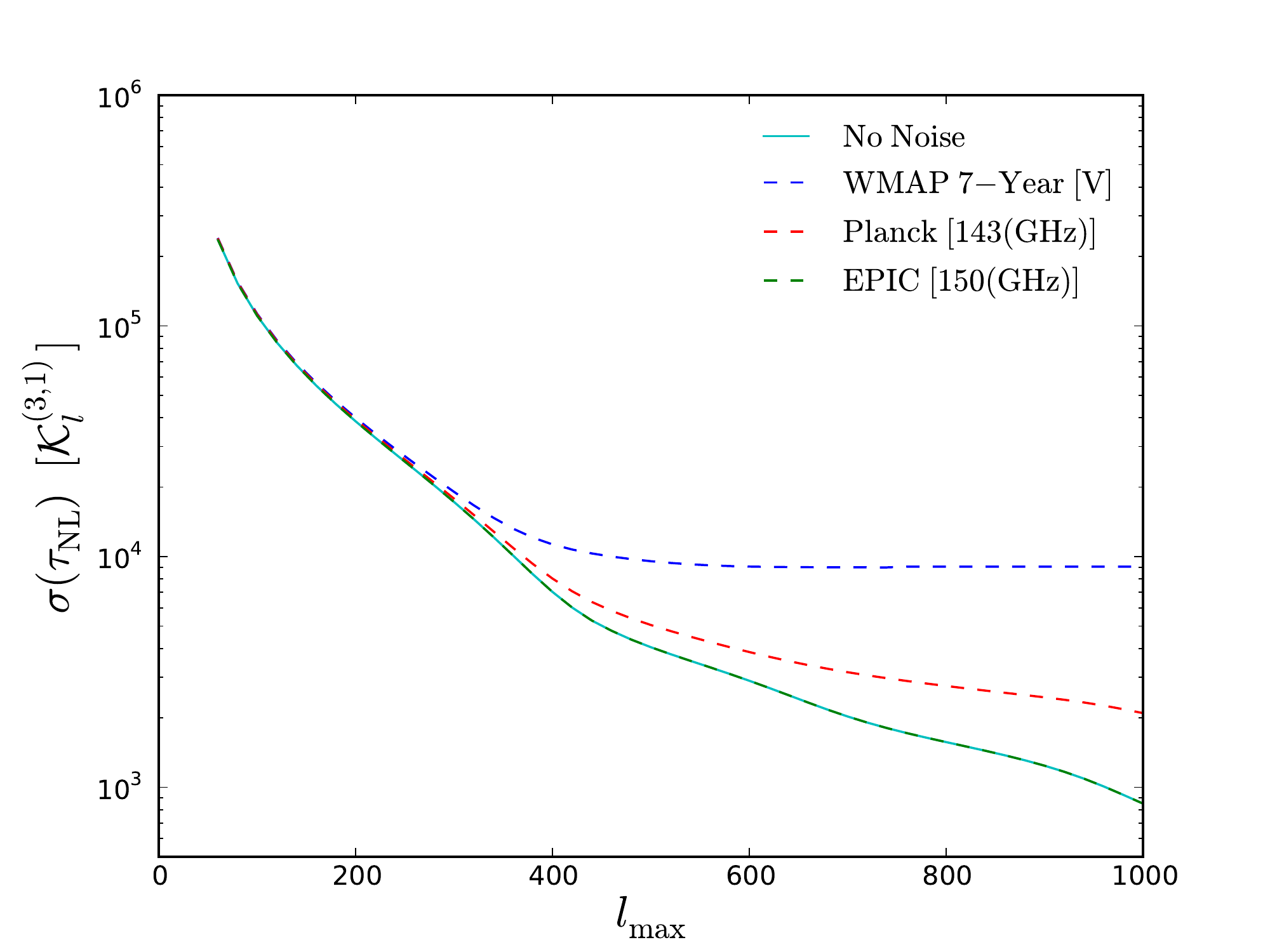}}
   \end{center}
   \vspace{-0.7cm}
   \caption[width=3in]{On top we have Fisher bounds on $\tau_{\rm NL}$ for the ${\cal K}_l^{(2,2)}$ estimator and on bottom for ${\cal K}_l^{(3,1)}$.  The frequency band used for each experiment is in brackets.}
   \label{fig:taunl}
\end{figure}
\begin{figure}
    \begin{center}
     {\includegraphics[scale=0.45]{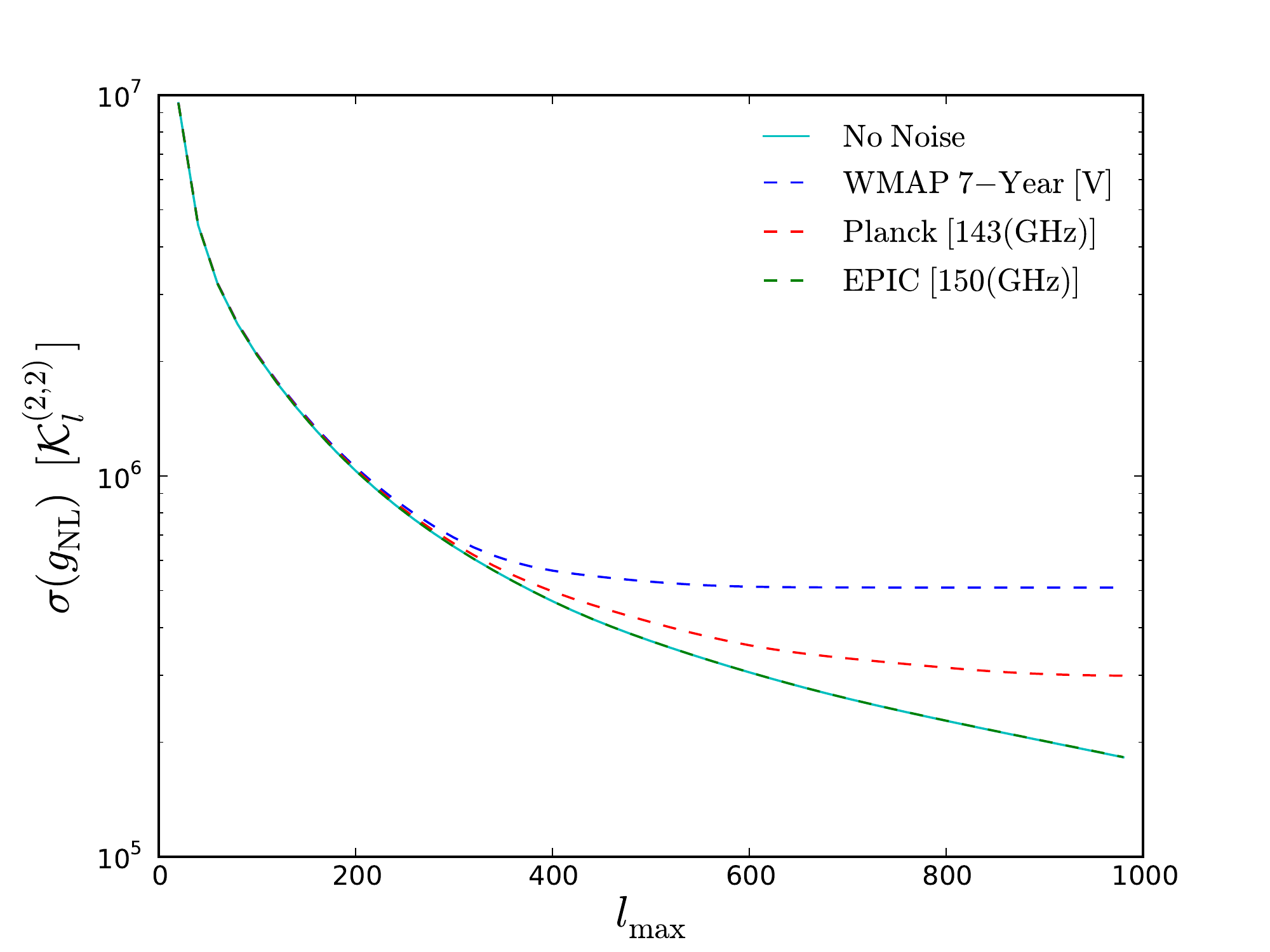} 
      \includegraphics[scale=0.45]{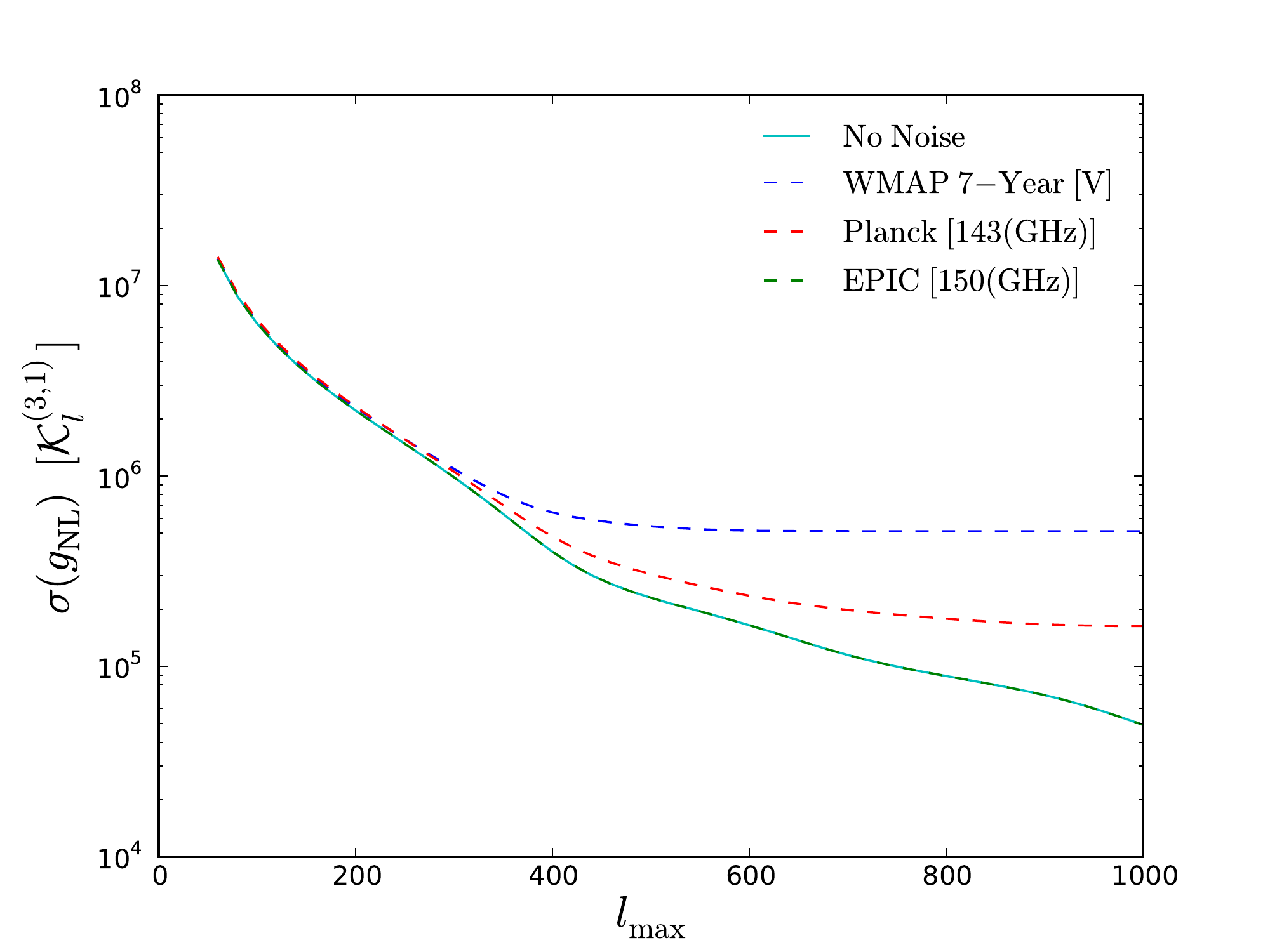}}
   \end{center}
   \vspace{-0.7cm}
   \caption[width=3in]{Top: Fisher bounds on $g_{\rm NL}$ for the ${\cal K}_l^{(2,2)}$ estimator. Bottom: Fisher bounds on $g_{\rm NL}$ for the ${\cal K}_l^{(3,1)}$ estimator.  The frequency band used for each experiment is in brackets.}
   \label{fig:gnl}
\end{figure}
Results from this estimator for $\tau_{\rm NL}$ are seen in
Fig.~\ref{fig:taunl}.  As with the $C_l^{(2,1)}$ estimator above, we show the
1$\sigma$ bound for the case without noise and beam as well as for the WMAP
7-Year, Planck and EPIC experiments.  For the case of no noise or beam, the
error bars for this estimator fall off as $\sim 1/\sqrt{\tau_{\rm NL} l^4}$.

Also plotted in the figure is the amplitude $A_{\rm NL}$ assuming $f_{\rm NL} =
32$, the WMAP 7-year best fit value.  Therefore, if $f_{\rm NL} = 32$ than we
must have  $A_{\rm NL} > 1$ for Planck to be able to have a detection of
$\tau_{\rm NL}$.  However, even if $A_{\rm NL} \sim 1$, EPIC should be able to
detect $\tau_{\rm NL}$, especially since EPIC will be able to use data much
past $l_{\rm max} = 1000$.

We also compute error bars for $\tau_{\rm NL}$ from our second skewness power
spectrum estimator for the trispectrum ${\cal K}_l^{(3,1)}$ by first
calculating the signal-to-noise from Eq.~\ref{eq:k31e} and~\ref{eq:Fk31e} then
solving for $\sigma(\tau_{\rm NL})$ from the fisher matrix~\ref{eq:fisher}.
Results for this calculation are plotted in Fig.~\ref{fig:taunl}.  Along with
the 1$\sigma$ error bars for each experiment, is the amplitude $A_{\rm NL}$
assuming $f_{\rm NL} = 50$.  The purpose of setting the amplitude to this value
is to demonstrate that if $f_{\rm NL}$ is large enough, models with $A_{\rm NL}
< 1$ may be able to be tested by upcoming experiments, especially EPIC.  

In addition to $\tau_{\rm Nl}$, bounds can be put on $g_{\rm NL}$ from the two
before mentioned four point estimators.  To do this, we calculate the
estimators from eq.~\ref{eq:k22e} and~\ref{eq:k31e} setting $\tau_{\rm NL} = 0$
and $g_{\rm NL} = 1$.  From here, we calculate the signal-to-noise from
eq.~\ref{eq:Fk22e} and~\ref{eq:Fk31e} whereupon we compute Fisher bounds from
equation~\ref{eq:fisher}.  The results are seen in Fig.~\ref{fig:gnl}.

\begin{figure}
    \begin{center}
      {\includegraphics[scale=0.45]{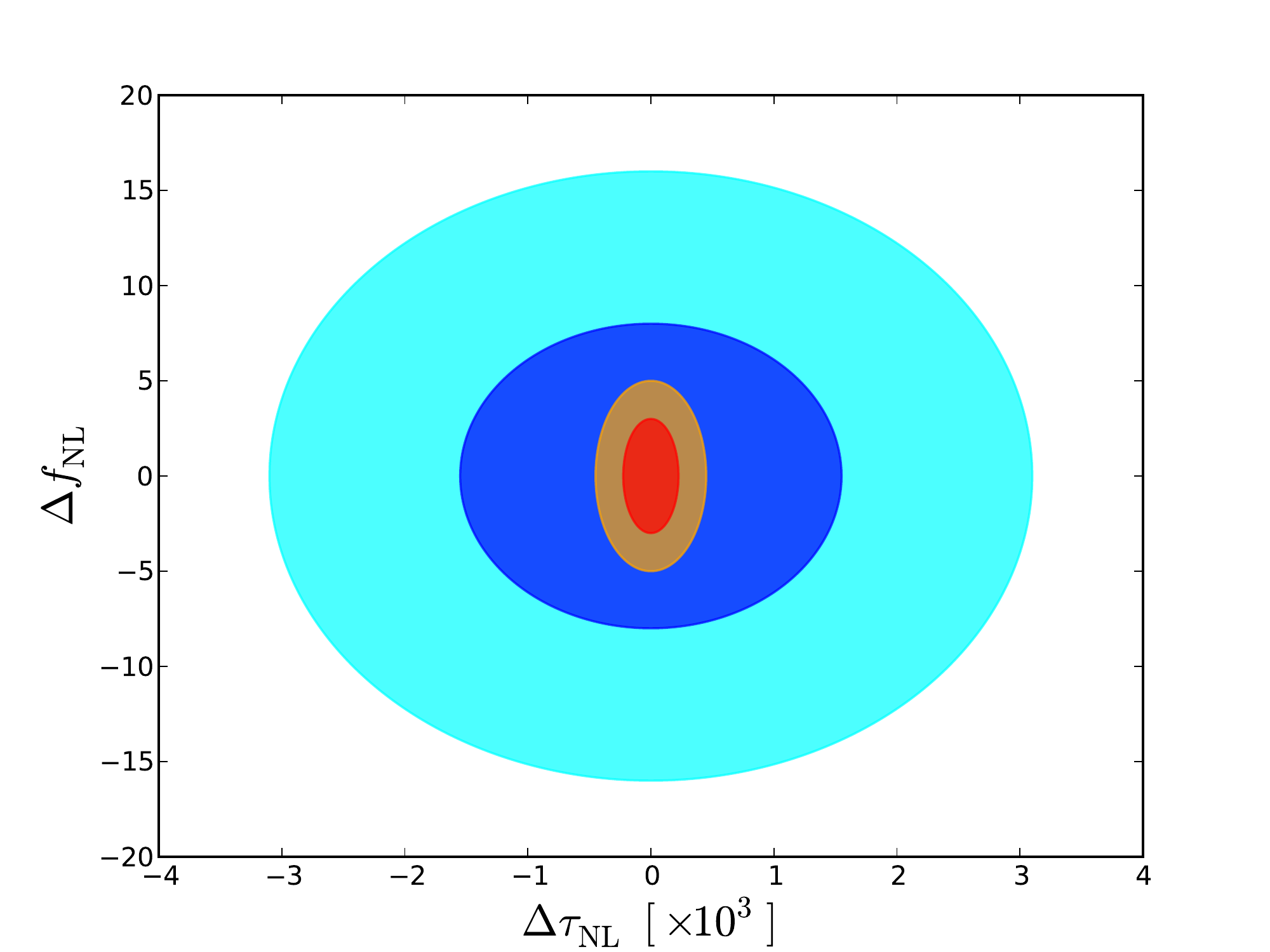} 
      \includegraphics[scale=0.45]{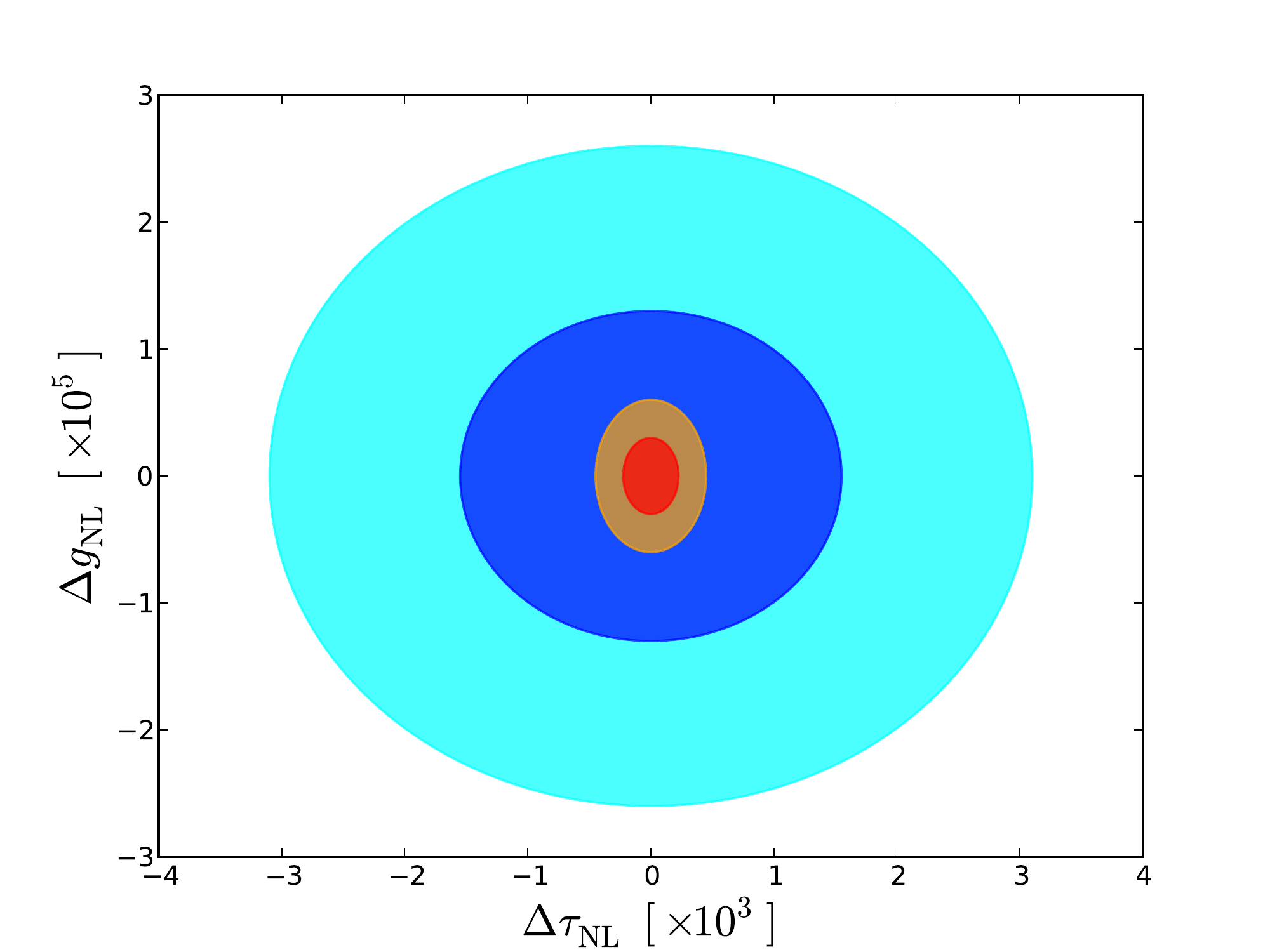}}
   \end{center}
   \vspace{-0.7cm}
   \caption[width=3in]{Fisher confidence intervals for $f_{\rm NL}$, $g_{\rm NL}$ and $\tau_{\rm NL}$.   The dark and light blue represent the 68\% and 95\% intervals respectively for Planck. The red and orange represent the 68\% and 95\% intervals respectively for EPIC.}
   \label{fig:epic}
\end{figure}

Combining the two estimators ${\cal K}_l^{(2,2)}$ and ${\cal K}_l^{(3,1)}$
gives the minimum error bars for $\tau_{\rm NL}$ and $g_{\rm NL}$ seen in
Table~\ref{tab:sigmabars} as well as Figure~\ref{fig:epic}.   These are
comparable to those of~\cite{Kogo:2006kh} and~\cite{Okamoto:2002ik} who
calculated Fisher bounds assuming only cosmic variance limited sky.   They did
not use the power skewness estimator, however, their estimator is equivalent in
the limit of homogeneous noise~\cite{Komatsu:2010fb}.  Kogo and Komastu~\cite{Kogo:2006kh} 
found a higher signal-to-noise than did Okamoto and Hu~\cite{Okamoto:2002ik}.  This paper finds a signal-to-nose in between these values.

\begin{table}[htbp]
  \centering
  \begin{tabular}{@{} |c|c|c|c|c| @{}}
\hline
    $l_{\rm max}$ & 500 & 1000 & 1500 & 2000\\ 
    \hline
        $f_{\rm NL}$ Planck & 16 & 10 & 8 & 8 \\ 
    \ \ \ \ EPIC & 15 & 7.5 & 5 & 3\\ 
    \hline
        $\tau_{\rm NL}$ Planck & 4350 & 1640 & 1550 & 1550 \\ 
    \ \ \ \ EPIC & 3700 & 920 & 400 & 225 \\ 
        \hline
             $g_{\rm NL}$ Planck & $1.6 \times 10^5 $ & $1.4 \times 10^5$ &  $1.3 \times 10^5$  & $1.3 \times 10^5$\\ 
    \ \ \ \ EPIC & $1.5 \times 10^5 $ & $1.1\times 10^5$ &  $8.4 \times 10^4$ & $6.0 \times 10^4$\\ 
            \hline
             $A_{\rm NL}$ Planck & $ 3.0 $ & $1.1 $ &  $1.0 $  & $1.0$\\ 
    \ \ \ \ EPIC & $2.5 $ & $0.6$ &  $0.3$ &0.15 \\ 
    \hline
      \end{tabular}
  \caption{The minimum error bars at 1$\sigma$ for $f_{\rm NL}$, $\tau_{\rm NL}$ and $g_{\rm NL}$, using both ${\cal K}_l^{(2,2)}$ and ${\cal K}_l^{(3,1)}$ estimators, for the Planck and EPIC experiments up to $l_{\rm max} = 2000$.  As stated in text, this assumes only one temperature frequency band is used in the analysis.}
  \label{tab:sigmabars}
\end{table}  

From this table we see that $\tau_{\rm NL}$ can be detected at $95\%$
confidence level by Planck if $\tau_{\rm NL} > 3000$ and EPIC for $\tau_{\rm
NL} > 600$.  If $f_{\rm NL} = 32$ in the bispectrum, this equivalently  means
$\tau_{\rm NL}$ can be detected if $A_{\rm NL} > 2$ and $A_{\rm NL} > 0.4$
respectively, again alluding to the fact that EPIC will be able to test some
inflationary models with $A_{\rm NL} < 1$.

\begin{figure}
    \begin{center}
      \includegraphics[scale=0.45]{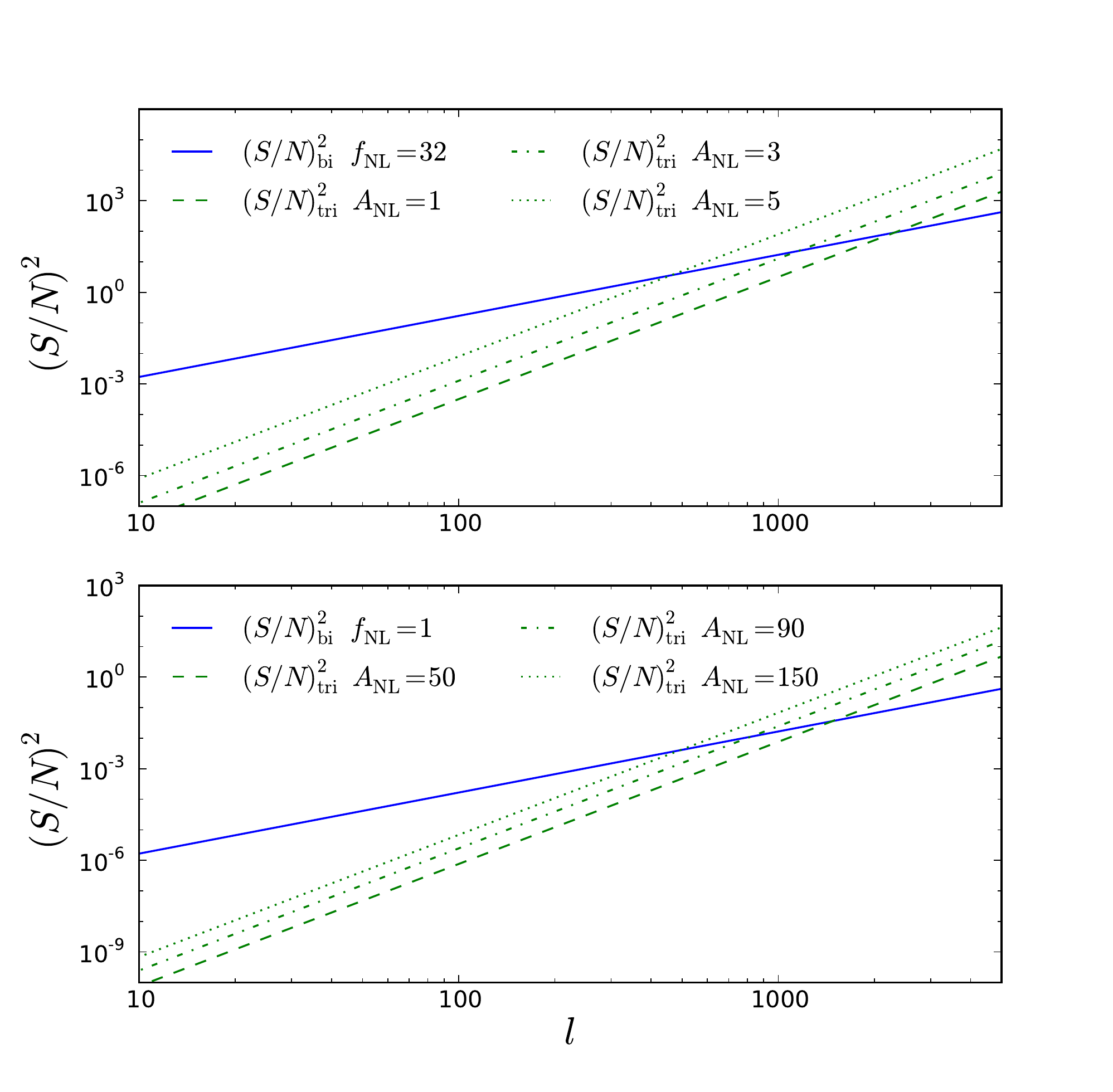} 
   \end{center}
   \vspace{-0.7cm}
   \caption[width=3in]{Comparison of the sensitivity of both the bispectrum and the trispectrum to non-Gaussianity assuming different values of $f_{\rm NL}$ and $A_{\rm NL}$.  }
   \label{fig:compare_f2A}
\end{figure}

Furthermore, as can be seen in Fig.~\ref{fig:compare_f2A}, for large enough
$A_{\rm NL}$, the trispectrum is more sensitive to non-Gaussianity, even for
Planck.  This may turn out to be very important as some models predict $A_{\rm
NL} > 1$.  It is therefore imperative that Planck examines the trispectrum for
non-Gaussianity as it may turn out to be more likely to get a detection there
than in the bispectrum.

Some models predict an undetectable amount of non-Gaussianity in the bispectrum
(For example, $f_{\rm NL} \sim 1$) with a large amount of non-Gaussianity in
the trispectrum.  These plots let us know just how {\it big} $A_{\rm NL}$ must
be in order for a detection of non-Gaussianity to be made in the trispectrum
for such scenarios. 

From these plots we see, for $f_{\rm NL} = 1$, the trispectrum becomes more
sensitive to non-Gaussianity than the bispectrum at $l = 1450$, $830$, and
$500$ for $A_{\rm NL} = 50$, 90 and 120 respectively.   For $f_{\rm NL} = 32$,
the trispectrum has more sensitivity at $l = 2350$, $1150$, and $450$ for
$A_{\rm NL} = 1$, 3 and 10 respectively and for  $f_{\rm NL} = 50$ we have more
sensitivity at $l = 1500$, $750$, and $300$ for $A_{\rm NL} = 1$, 3 and 10
respectively. 

\begin{figure}
    \begin{center}
      {\includegraphics[scale=0.35]{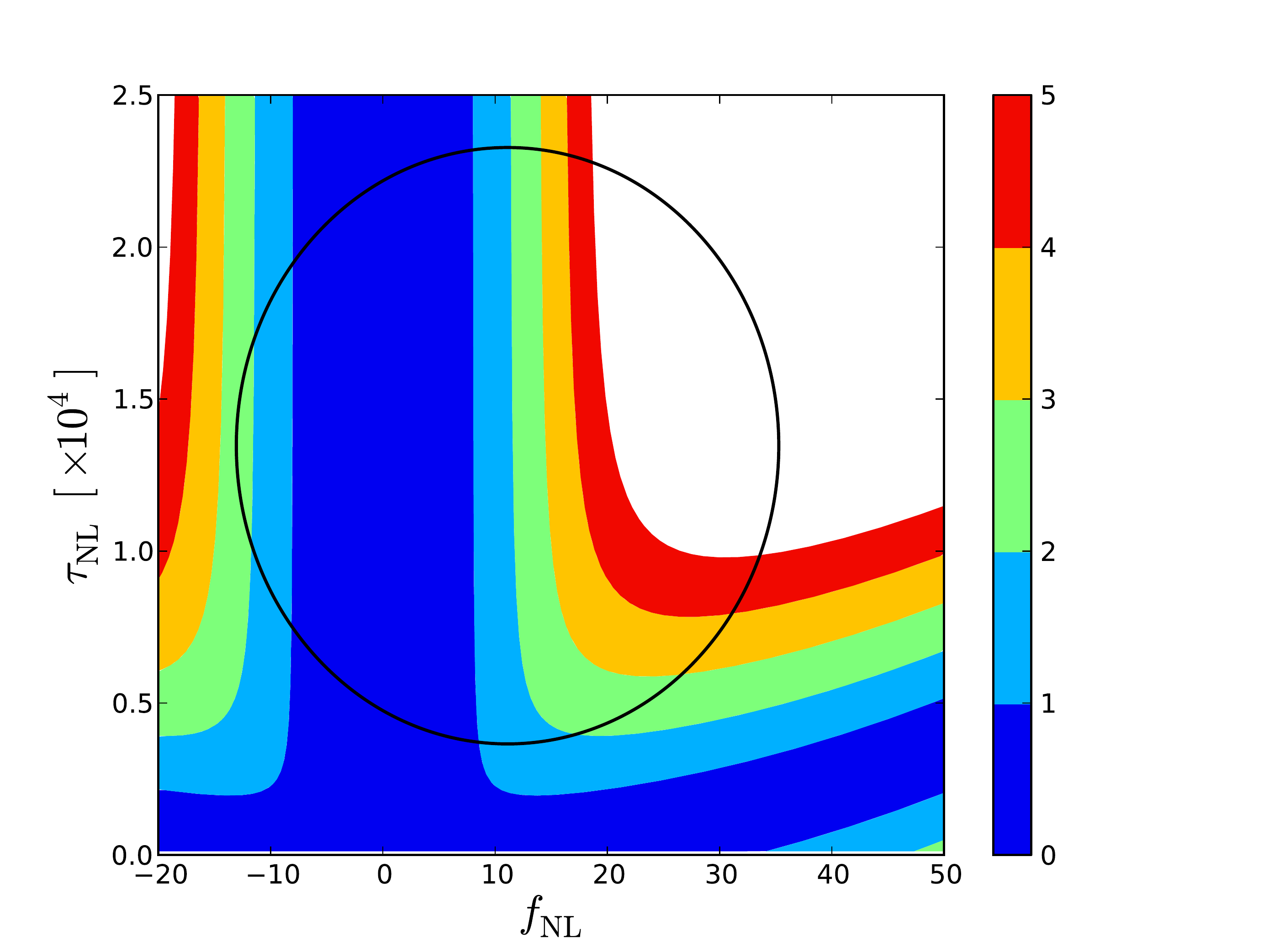} 
      \includegraphics[scale=0.35]{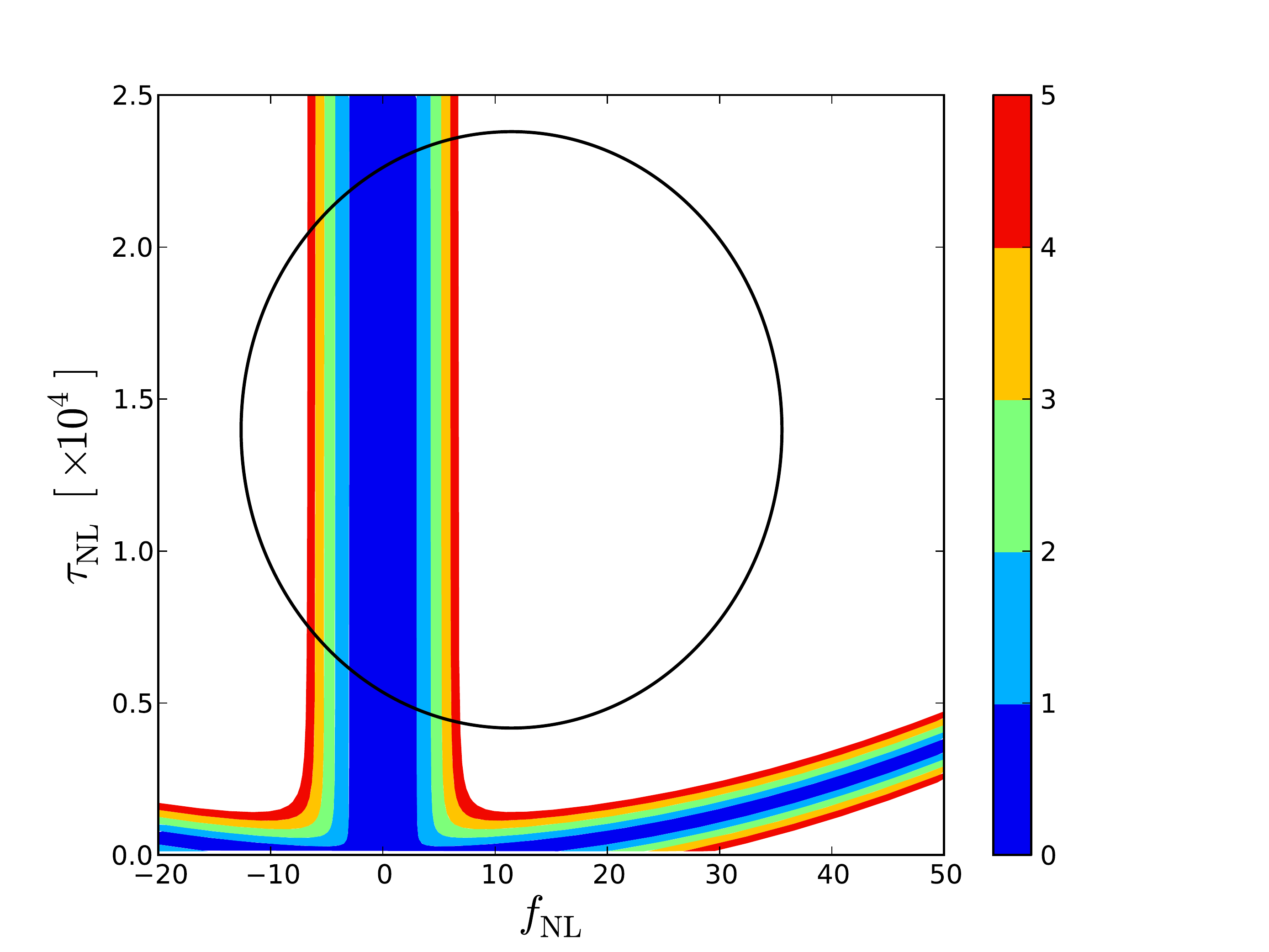} }
   \end{center}
   \vspace{-0.7cm}
   \caption[width=3in]{The top plot shows $(A_{\rm NL} - 1)/\Delta A_{\rm NL}$  for Planck and the bottom for EPIC.  The color bands show to how many sigma $A_{\rm NL}$ would differ from unity for possible best fit values for Planck and EPIC.  If Planck or EPIC find best fit $f_{\rm NL}$ and $\tau_{\rm NL}$ values anywhere in the white region, single-field inflation will be ruled out by $> 5\sigma$.  The black ellipse marks the 68\% confidence region for the Smidt et al.(2009) best fit $f_{\rm NL}$ and Smidt et al.(2010) best fit $\tau_{\rm NL}$ values respectively~\cite{Smidt:2009ir}.} 
   \label{fig:danl}
\end{figure}

Figure~\ref{fig:danl} shows $(A_{\rm NL} - 1)/\Delta A_{\rm NL}$ for both
Planck and EPIC.  In this plot it is clear that both Planck and EPIC are in a
position to rule out single field inflation by determining $A_{\rm NL} \neq 1$.
Large sections of the parameter space, consistent with current measurements,
will rule out $A_{\rm NL}$ equal to unity by $> 5\sigma$.  

Note from table \ref{tab:sigmabars} that the expected bound on $g_{\rm NL}$ is about two orders of magnitude weaker than that on $\tau_{\rm NL}$, even though both parameters are suppressed by a power spectrum cubed in (\ref{taug}). One reason is that the $k$ dependent shape factor multiplying $\tau_{\rm NL}$ in (\ref{taug}) diverges whenever $k_{ij}\rightarrow0$, while the factor multiplying $g_{\rm NL}$ only diverges when one of the $k_i\rightarrow0$ (and in this case the same applies for $\tau_{\rm NL}$ as well).

%
%

\section{Prior Analysis Using These Estimators}
\label{sec:prioranalysis}

These skewness and kurtosis power spectrum estimators have recently been
employed to constrain non-Gaussianity in the WMAP 5-year data.  Using the
bispectrum, Smidt et al. (2009) found that $-36.4 < f_{\rm NL} < 58.4$ at 95\%
confidence~\cite{Smidt:2009ir}.  This bound puts the 1$\sigma$ error bars at
$\pm 23.5$, within about 12\% of the optimal Fisher bound.

The analysis for the trispectrum is more difficult and we therefore elaborate about it here.  Our recipe for analysis is
\begin{enumerate}
\item We calculate ${\cal K}_l^{(3,1)} $ and ${\cal K}_l^{(2,2)}$ in Eq.~\ref{eq:k22e}-~\ref{eq:k31e} for $\tau_{\rm NL}$ and $g_{\rm NL} = 1$. 
\item We extract ${\cal K}_l^{(3,1)} $ and ${\cal K}_l^{(2,2)}$ directly from WMAP 5-year data.
\item We perform the extraction of ${\cal K}_l^{(3,1)} $ and ${\cal K}_l^{(2,2)}$ from 250 Gaussian maps, allowing us to determine error bars and the Gaussian piece of each estimator.
\item We subtract off the Gaussian contribution to these estimators to ensure we are fitting to the non-Gaussian contribution.
\item We fit the two unknowns $\tau_{\rm NL}$ and $g_{\rm NL}$ from data using the two equations simultaneously.  The amplitudes the theoretical curves must be scaled by gives the values for $\tau_{\rm NL}$ and $g_{\rm NL}$
\item We constrain $A_{\rm NL}$ by comparing $\tau_{\rm NL}$ from the trispectrum with $(6f_{\rm NL}/5)^2$ coming from the bispectrum.
\end{enumerate}
This recipe is described in grater detail below:

First we calculate  ${\cal K}_l^{(3,1)} $ and ${\cal K}_l^{(2,2)}$ theoretically using equations Eq.~\ref{eq:reducedtri}-~\ref{eq:FL} and Eq.~\ref{eq:k22e}-~\ref{eq:k31e} for a model with $\tau_{\rm NL}$ and $g_{\rm NL} = 1$.  To obtain $C_l$ we use CAMB~\cite{Lewis:1999bs}\footnote{${\rm http://camb.info/}$} with the WMAP 5-year best fit parameters and use the beam transfer functions from the WMAP team.    We then obtain the connected piece using a modified version of the CMBFAST code~\cite{Seljak:1996is}\footnote{${\rm http://www.cfa.harvard.edu/~mzaldarr/CMBFAST/cmbfast.html}$}. Plots of many of the quantities used for these calculations can be found in Ref.~\cite{Smidt:2009ir}. 

\begin{figure}[t]
   \vspace{-0.2cm}
    \begin{center}
      \includegraphics[scale=0.45]{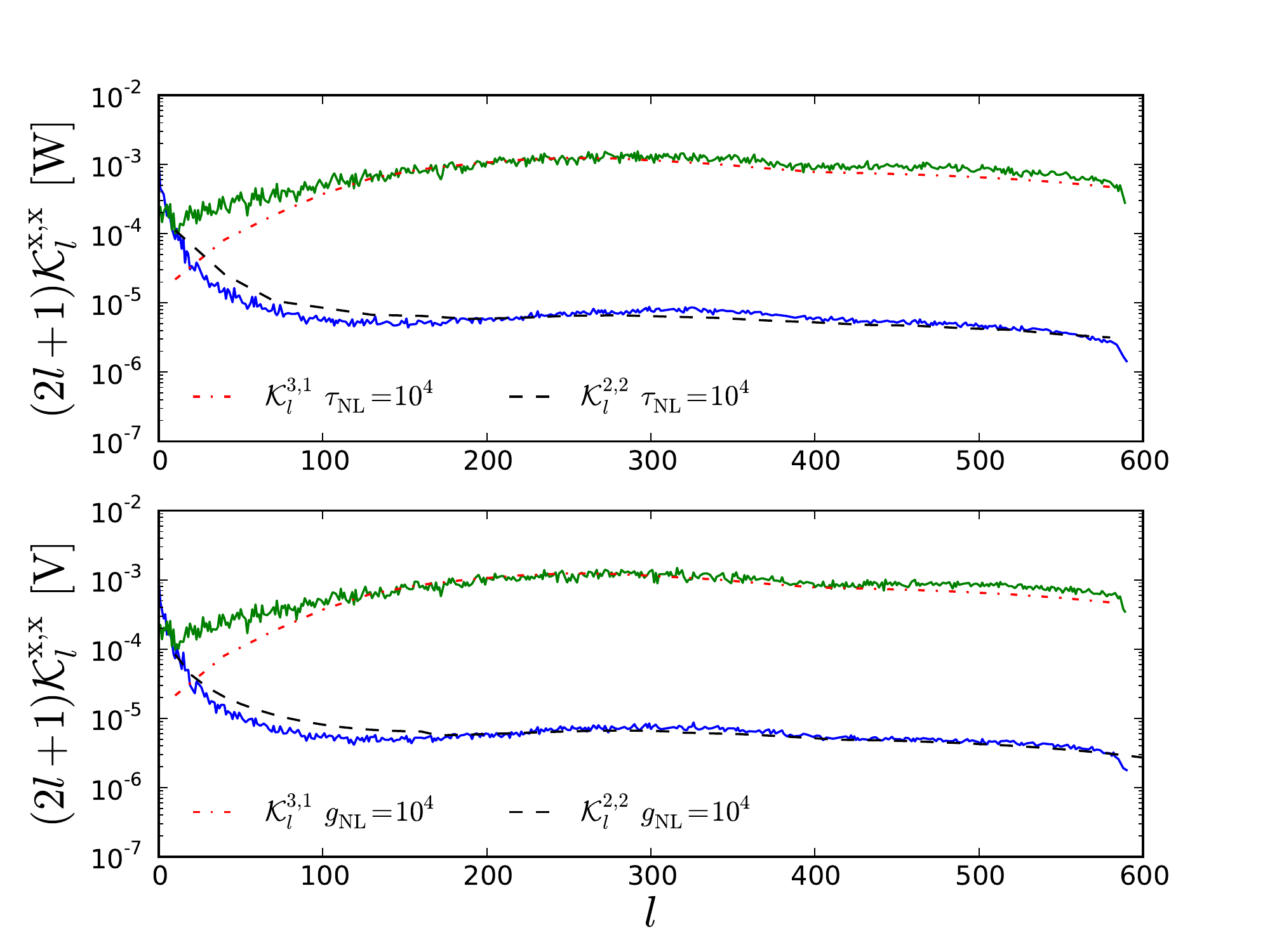} 
   \end{center}
   \vspace{-0.7cm}
   \caption[width=3in]{The top plot shows the ${\cal K}_l^{3,1}$ and ${\cal K}_l^{2,2}$ estimators, shown in green and blue respectively, taken from data for the W band.  The same estimators for the V band are shown on the bottom.  Additionally on the top the theoretical contributions for  ${\cal K}_l^{2,2}$ and ${\cal K}_l^{3,1}$ proportional to $\tau_{\rm NL}$ are shown with the bottom showing those proportional to $g_{\rm NL}$.  The Gaussian contributions were not removed from these plots.\vspace{-0.1cm}}
   \label{fig:wtheory}
\end{figure}

Plots of ${\cal K}_l^{(2,2)} $ and ${\cal K}_l^{(3,1)}$ are shown in Figure~\ref{fig:wtheory}. These curves will be compared with estimators derived from data to determine the magnitude of each statistic.  Since we have two estimators, we can solve for the two unknowns $\tau_{\rm NL}$ and $g_{\rm NL}$ by fitting both estimators simultaneously.  

To calculate\footnote{see Smidt el al. 2009 for a similar calculation using the bispectrum for more details.~\cite{Smidt:2009ir}} the estimators from data, used in the lefthand side of equations~(\ref{eq:k22e}) and~(\ref{eq:k31e}), we use both the raw and foreground-cleaned WMAP 5-Year Stokes I maps for V- and W-bands masked with the KQ75 mask~\footnote{${\rm http://lambda.gsfc.nasa.gov/}$}. We use the Healpix library to analyze the maps.  For this analysis we only considered data out to $l_{\rm max} = 600$.  We correct for the KQ75 mask using a matrix $M_{l l'}$, based on the power spectrum of the mask, as described above.

Figure~\ref{fig:wtheory} shows the results for ${\cal K}_l^{3,1}$ and ${\cal K}_l^{2,2}$ for the V and W frequency bands extracted from the raw WMAP 5-Year maps. 
In order to do proper statistics for our data fitting we create 250 simulated Gaussian maps of each frequency band with $n_{\rm side}=512$.  To obtain Gaussian maps we run the {\it synfast} routine of Healpix with an in-file representing the WMAP 5-year best-fit CMB anisotropy power spectrum and generate maps with information out to $l = 600$.  We then use {\it anafast}, without employing an iteration scheme, masking with the $KQ75$ mask, to produce $a_{lm}$'s for the Gaussian maps out to $l=600$.  Obtaining estimators from these Gaussian maps allows us to uncover the Gaussian contribution to each estimator in addition to providing us information needed to calculate the error bars on our results.  

This whole process is computationally intensive.  To calculate all theoretical estimators took nearly 8,000 CPU hours.  Furthermore, all the estimators from Gaussian and data maps combined took an additional 1600 CPU hours.
\begin{figure}
   \vspace{-0.2cm}
    \begin{center}
      \includegraphics[scale=0.45]{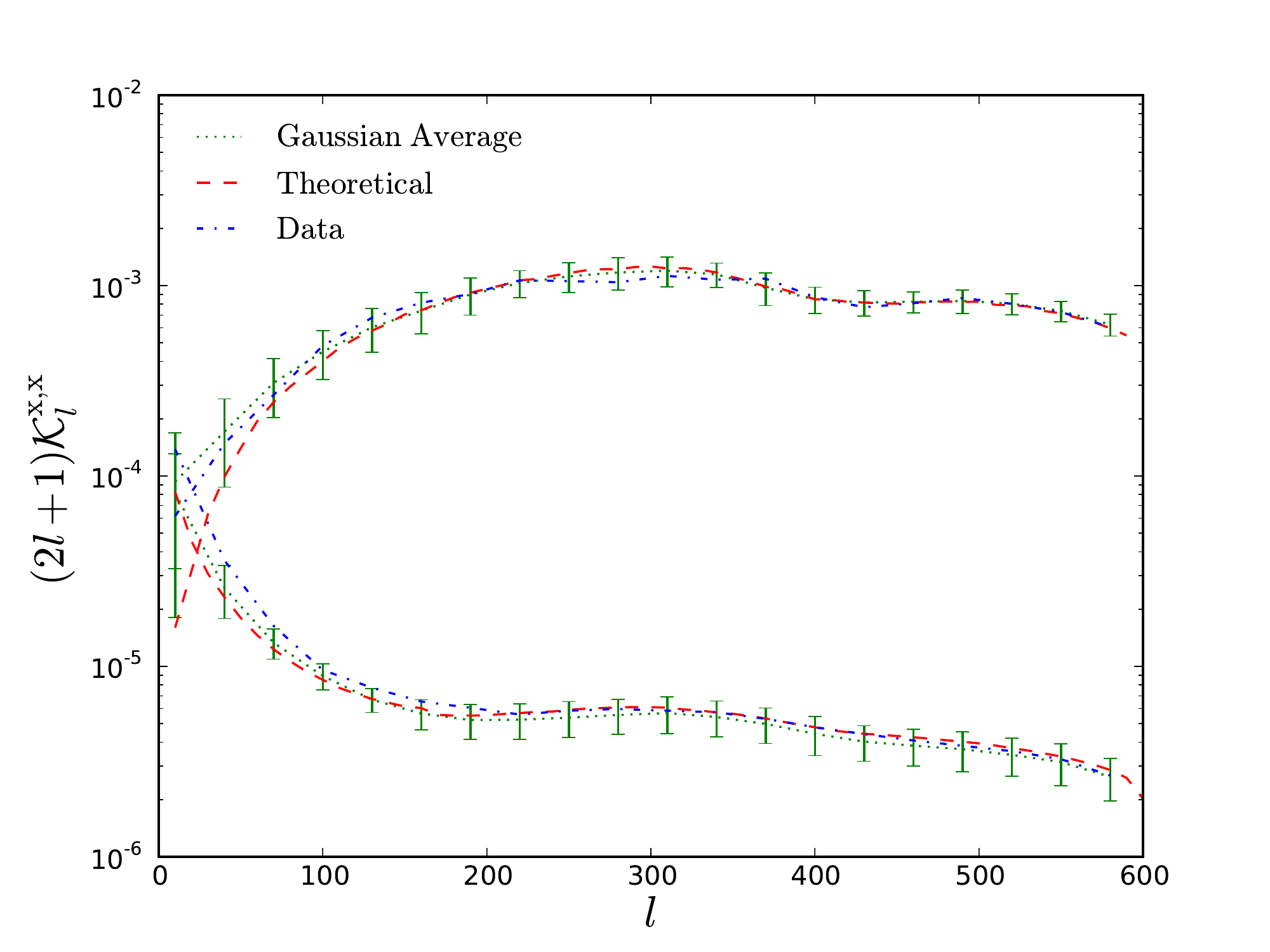} 
   \end{center}
   \vspace{-0.5cm}
   \caption[width=3in]{The relation between the full estimators coming from data versus the Gaussian contributions.  The green curve show the Gaussian contributions coming from averaging the estimators from the Gaussian maps.  The red curve is the theoretical Gaussian piece calculated using the WMAP-5 best-fit cosmology power spectrum.  The error bars show two standard deviations from the Gaussian curves.  These curves are from W band data.\vspace{-0.1cm}}
   \label{fig:compare}
\end{figure}
\begin{figure}[t]
   \vspace{-0.2cm}
    \begin{center}
      \includegraphics[scale=0.48]{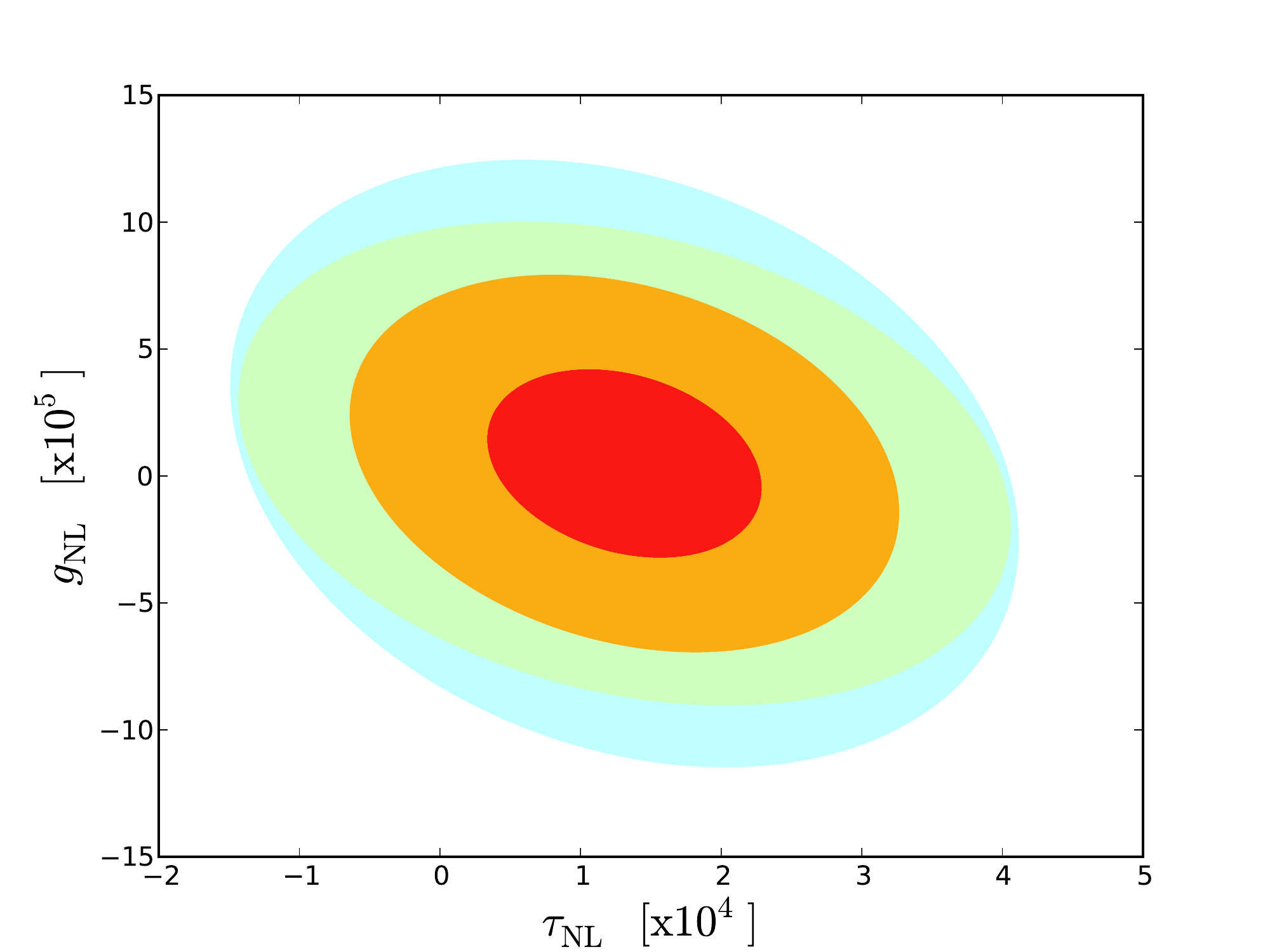} 
   \end{center}
   \vspace{-0.7cm}
   \caption[width=3in]{The 95\% confidence levels for $g_{\rm NL}$ versus $\tau_{\rm NL}$.  The red and orange represent the 68\% and 95\% intervals respectively for the combined V+W analysis.  The light blue regions represent the 95\% confidence intervals for the V band analysis, and the light green regions are for the W band.}
   \label{fig:confidence}
\end{figure}

As previously discussed, the full trispectrum can be decomposed into both a Gaussian and non-Gaussian or connected piece.  To make a measurement of  non-Gaussianity we to subtract off the Gaussian piece from the full trispectrum.  Figure~\ref{fig:compare} shows the the relationship between the full trispectrum and the Gaussian piece.  In this plot the Gaussian piece was calculated in two different ways as a sanity check.  First, the Gaussian maps were averaged over.  Second, the Gaussian piece of each estimator is calculated theoretically using Eq.~\ref{eq:gausspiece}.  

\begin{center}
\begin{table*}[ht]
{\small
\hfill{}
\begin{tabular}{|c|c|c|c|}
\hline
Band & W & V & V+W \\
\cline{3-4}
\hline
Raw & & & \\
$g_{\rm NL}$  & $4.7 \times 10^4 \pm 5.3  \times 10^5$ & $4.6 {\rm x} 10^4 \pm 5.9  \times 10^5$& $4.7 {\rm x} 10^4 \pm 3.9  \times 10^5$ \\ 
    $\tau_{\rm NL}$ & ${(1.63 \pm 1.27)}  \times 10^4$ &  ${(1.68 \pm 1.31)}  \times 10^4$ &  ${(1.64 \pm 0.98)}  \times 10^4$  \\ 
        $A_{\rm NL}$ &  $7.4 \pm 7.3$& $6.3 \pm 6.0$  &  $11.1 \pm 7.3$  \\ 
       \hline
       FC & & & \\
        $g_{\rm NL}$  & $4.2 \times 10^4 \pm 5.3  \times 10^5$ & $4.1 \times 10^4 \pm 5.9  \times 10^5$& $4.2 \times10^4 \pm 3.9  \times 10^5$ \\ 
    $\tau_{\rm NL}$ &  ${(1.32 \pm 1.27)}  \times 10^4$ &   ${(1.39 \pm 1.31)}  \times 10^4$ &  ${(1.35 \pm 0.98)}  \times10^4$  \\ 
            $A_{\rm NL}$ &  $6.0 \pm 6.7$& $5.2 \pm 5.7$  & $9.2 \pm 6.1$  \\ 
\hline
\end{tabular}}
\hfill{}
\caption[width=0.55]{Results for each frequency band to $1\sigma$.  Values for $g_{\rm NL}$, $\tau_{\rm NL}$ and $A_{\rm NL}$ on the top are for raw maps.  The values on the bottom are for foreground clean maps. $A_{\rm NL}$ is estimated assuming $f_{\rm NL}=32 \pm 21$ from the WMAP-7 analysis and the tabulated 1$\sigma$ uncertainy 
is based on an analytical error propagation.
\vspace{-0.3cm}}
\label{tab:label}
\end{table*}
\end{center}

After obtaining the theory, data and simulated curves we use the best fitting procedure described in~\cite{Smidt:2009ir} where we minimize $\chi^2$ to fit $\tau_{\rm NL}$ and $g_{\rm NL}$ simultaneously.  Our results are listed in Table~\ref{tab:label}.  We see that $g_{\rm NL}$ and $\tau_{\rm NL}$ are consistent with zero with 95\% confidence level ranges  $-7.4 < g_{\rm NL}/10^5 < 8.2$ and   $-0.6 < \tau_{\rm NL}/10^4 < 3.3$ for V+W-band in foreground-cleaned maps.  The 95\% confidence intervals of $g_{\rm NL}$ versus $\tau_{\rm NL}$ are plotted in Figure~\ref{fig:confidence} for each band.   For a V band analysis alone, the 68\% confidence intervals
 are $\tau_{\rm NL} = (1.39 \pm 1.31)\times10^4$ and 
 $g_{\rm NL} =  4.6\times10^4 \pm 5.9\times10^5$.
These error bars are within $\sim$40\% and $\sim$20\% of the optimal Fisher values
discussed above comparing with WMAP 7-year level noise for $\tau_{\rm NL}$
and $g_{\rm NL}$ respectively.

Combining $f_{\rm NL}=11 \pm 24$ from Ref.~\cite{Smidt:2009ir} and $\tau_{\rm
NL}= (1.35 \pm 0.98) \times 10^4$ from our skewness analysis we get $-649 <
A_{\rm NL} < 805$ at 95\% confidence.  If instead we had assumed $f_{\rm NL}=32
\pm 21$ from WMAP-7 analysis~\cite{Komatsu:2010fb} and same $\tau_{\rm NL}$
reported here we find $-3 < A_{\rm NL} < 21.4$ at 95\%
confidence. The difference of the two estimates is a reflection on the central
value of $f_{\rm NL}$ since $A_{\rm NL}$ = $\tau_{\rm NL}/(6f_{\rm NL}/5)^2$
and therefore a smaller $f_{\rm NL}$ results in a larger uncertainty in $A_{\rm
NL}$.  This behavior is also seen in Fig.~\ref{fig:danl}.

No measurements involving WMAP 7-year data have been preformed using these
estimators.  It is our opinion that the results for WMAP 7-year data will not
be much different than for the WMAP 5-year data, just as the optimal results
using the traditional skewness statistic $S_3$ do not differ significantly
between these two data sets.~\cite{Smith:2009jr,Komatsu:2010fb}

Planck, on the other hand, is in a position to make significant improvements in
the measurement of non-Gaussianity using these estimators.  Since Planck is
taking data, we encourage any plans to measure $f_{\rm NL}$, $g_{\rm NL}$ and
$\tau_{\rm NL}$ using the skewness and kurtosis spectrum statistics that we
have proposed.  In addition to ruling out the standard single-field slow-roll
inflation model with a detection of non-Gaussianity in general, Planck is in a
position to possibly rule out all single-field models with a measurement of
$A_{\rm NL} \neq 1$.

\section{Conclusions} 
\label{sec:conclusion}

In this paper we discussed the skewness and kurtosis power spectrum approach to
probing primordial non-Gaussianity.  We outlined the expected constraints these
techniques will place using future experimental data.  These constraints were
calculated by computing the signal-to-noise for each estimator, properly taking
into account the noise and beam of each experiment.  Optimal error bars for
$f_{\rm NL}$,  $g_{\rm NL}$ and $\tau_{\rm NL}$ are listed as a function of
$l_{\rm max}$.  

It was argued that the skewness and kurtosis power spectrum approach to measure
non-Gaussianity has several advantages. These advantages include the ability to
separate foregrounds and other secondary non-Gaussian signals, the ability to
measure the scale dependance of each statistic and an advantage that the cut
sky can be corrected from a matrix $M_{l l'}$ without needing to compute extra
linear terms.  

The physical significance of each non-Gaussian statistic is discussed.  In the
bispectrum, different non-Gaussian triangle configurations in Fourier space
contributing to $f_{\rm NL}$ are related to different underlying physics.  By
adding a local measurement of the trispectrum, a new statistic $A_{\rm NL} =
\tau_{\rm NL}/(6 f_{\rm NL}/5)^2$ will be a powerful probe to  distinguish
between multi-field models.  Single-field models can be ruled out in general if
$A_{\rm NL} \neq 1$ and we discussed how this may be a real possibility with
Planck or EPIC.   Furthermore, for $A_{\rm NL}$ large enough, the trispectrum
becomes a better probe for non-Gaussinity than the bispectrum for analysis
utilizing information on very small scales. The parameter $g_{\rm NL}$ will be
the hardest to constrain.  A constraint on this parameter will uncover
information on self-interactions. 

\acknowledgments
We are grateful to Eiichiro Komatsu, Tomo Takahashi and Paolo Serra for assistance during
various stages of this work.  This work was supported by NSF CAREER AST-0645427
and NASA NNX10AD42G at UCI and STFC rolling grant ST/G002231/1
(DM).\vspace{-0.4cm}

\appendix

%
%

\end{document}